\documentclass{jpp-AAS}

\usepackage{graphicx}
\usepackage{natbib}

\usepackage{amsmath}
\usepackage{amssymb}
\usepackage[makeroom]{cancel}
\usepackage{color}
\usepackage[colorlinks=true,linkcolor=blue,citecolor=blue]{hyperref}

\ifCUPmtlplainloaded \else
  \checkfont{eurm10}
  \iffontfound
    \IfFileExists{upmath.sty}
      {\typeout{^^JFound AMS Euler Roman fonts on the system,
                   using the 'upmath' package.^^J}%
       \usepackage{upmath}}
      {\typeout{^^JFound AMS Euler Roman fonts on the system, but you
                   dont seem to have the}%
       \typeout{'upmath' package installed. JPP.cls can take advantage
                 of these fonts, if you use 'upmath' package.^^J}%
      }
  \else
  \fi
\fi

\ifCUPmtlplainloaded \else
  \checkfont{msam10}
  \iffontfound
    \IfFileExists{amssymb.sty}
      {\typeout{^^JFound AMS Symbol fonts on the system, using the
                'amssymb' package.^^J}%
       \usepackage{amssymb}%
         
       \let\ge=\geqslant  
      }{}
  \fi
\fi

\ifCUPmtlplainloaded \else
  \IfFileExists{amsbsy.sty}
    {\typeout{^^JFound the 'amsbsy' package on the system, using it.^^J}%
     \usepackage{amsbsy}}
    {\providecommand\boldsymbol[1]{\mbox{\boldmath $##1$}}}
\fi

\newcommand{\figref}[1]{figure~\ref{#1}}

\newcommand{\Secref}[1]{Section~\ref{#1}}
\newcommand{\secref}[1]{section~\ref{#1}}

\newcommand{\secsand}[2]{sections~\ref{#1} and \ref{#2}}
\newcommand{\secsdash}[2]{sections~\ref{#1}--\ref{#2}}
\newcommand{\secsref}[1]{sections~\ref{#1}}

\newcommand{\Eqref}[1]{Equation~(\ref{#1})}
\renewcommand{\eqref}[1]{equation~(\ref{#1})}

\newcommand{\exref}[1]{(\ref{#1})}
\newcommand{\exsdash}[2]{(\ref{#1}--\ref{#2})}

\newcommand{\bea}{\begin{eqnarray}}
\newcommand{\eea}{\end{eqnarray}}
\newcommand{\beq}{\begin{equation}}
\newcommand{\eeq}{\end{equation}}
\newcommand{\lt}{\left}
\newcommand{\rt}{\right}

\newcommand{\la}{\langle}
\newcommand{\ra}{\rangle}

\newcommand{\mbf}[1]{\boldsymbol{#1}}

\newcommand{\hence}{\quad\Rightarrow\quad}

\newcommand{\rmwhere}{\quad\mathrm{where}\quad}

\newcommand{\dd}{\partial}
\newcommand{\vdel}{\mbf{\nabla}}
\newcommand{\vdperp}{\vdel_\perp}
\newcommand{\dperp}{\nabla_\perp}

\newcommand{\hdel}{\hat{\nabla}}

\newcommand{\hdperp}{\hdel_\perp}

\newcommand{\hvdperp}{\hat{\vdel}_\perp}
\newcommand{\dpar}{\nabla_\parallel}

\newcommand{\rmd}{\mathrm{d}}

\newcommand{\eps}{\varepsilon}
\newcommand{\epsA}{\eps_\mathrm{AW}}
\newcommand{\epsC}{\eps_\mathrm{compr}}
\newcommand{\epsK}{\eps_\mathrm{KAW}}
\newcommand{\epsI}{\eps_\mathrm{ICW}}
\newcommand{\Qx}{Q_\mathrm{x}}
\newcommand{\WA}{W_\mathrm{AW}}
\newcommand{\WS}{W_\mathrm{SW}}
\newcommand{\WK}{W_\mathrm{KAW}}
\newcommand{\WI}{W_\mathrm{ICW}}
\newcommand{\XK}{X_\mathrm{KAW}}
\newcommand{\XI}{X_\mathrm{ICW}}

\newcommand{\aext}{a_\mathrm{ext}}
\newcommand{\baext}{\overline{a}_\mathrm{ext}}
\newcommand{\taext}{\widetilde{a}_\mathrm{ext}}
\newcommand{\vjext}{\vj_\mathrm{ext}}
\newcommand{\jext}{j_{\parallel\mathrm{ext}}}
\newcommand{\Jext}{\mathcal{J}_\mathrm{ext}}
\newcommand{\bJext}{\overline{\mathcal{J}}_\mathrm{ext}}
\newcommand{\tJext}{\widetilde{\mathcal{J}}_\mathrm{ext}}
\newcommand{\ephi}{\varphi}
\newcommand{\bphi}{\overline{\ephi}}
\newcommand{\tphi}{\widetilde{\ephi}}
\newcommand{\Apar}{A_\parallel}
\newcommand{\vAA}{\mbf{A}}
\newcommand{\vAperp}{\mbf{A}_\perp}
\newcommand{\vj}{\mbf{j}}
\newcommand{\A}{\mathcal{A}}
\newcommand{\bA}{\overline{\A}}
\newcommand{\tA}{\widetilde{\A}}
\newcommand{\dvB}{\delta\!\vB}
\newcommand{\dB}{\delta\! B}
\newcommand{\dtB}{\delta\!\widetilde{B}}
\newcommand{\dbB}{\delta\!\overline{B}}
\newcommand{\dBperp}{\dB_\perp}
\newcommand{\dn}{\delta n}
\newcommand{\dtn}{\delta\widetilde{n}}
\newcommand{\dbn}{\delta\overline{n}}

\newcommand{\df}{\delta\! f}
\newcommand{\vR}{{\mbf{R}}}
\newcommand{\rhos}{\rho_\mathrm{s}}
\newcommand{\rhoh}{\rho_\mathrm{H}}
\newcommand{\cs}{c_\mathrm{s}}
\newcommand{\vs}{v_\mathrm{s}}
\newcommand{\vrho}{{\mbf{\rho}}}
\newcommand{\hvpar}{\hat v_\parallel}
\newcommand{\hvperp}{\hat v_\perp}
\newcommand{\hJ}{\hat J}
\newcommand{\hG}{\hat \Gamma}
\newcommand{\bh}{\overline{h}}
\newcommand{\bg}{\overline{g}}
\newcommand{\tg}{\widetilde{g}}
\newcommand{\G}{\mathcal{G}}
\newcommand{\tW}{\widetilde W}
\newcommand{\B}{\mathcal{B}}
\newcommand{\U}{\mathcal{U}}
\newcommand{\bPsi}{\overline{\Psi}}
\newcommand{\bPhi}{\overline{\Phi}}
\newcommand{\bB}{\overline{\B}}
\newcommand{\bU}{\overline{\U}}
\newcommand{\tPsi}{\widetilde{\Psi}}
\newcommand{\tPhi}{\widetilde{\Phi}}
\newcommand{\tB}{\widetilde{\B}}
\newcommand{\tU}{\widetilde{\U}}

\newcommand{\Z}{\mathcal{Z}}

\newcommand{\vr}{{\mbf{r}}}

\newcommand{\vk}{{\mbf{k}}}

\newcommand{\kperp}{k_\perp}
\newcommand{\kcoll}{k_{\perp\nu}}

\newcommand{\kpar}{k_\parallel}

\newcommand{\vv}{\mbf{v}}
\newcommand{\vvperp}{\vv_\perp}
\newcommand{\vperp}{v_\perp}
\newcommand{\dvperp}{\delta\vperp}
\newcommand{\vpar}{v_\parallel}
\newcommand{\vth}{v_{{\rm th}}}

\newcommand{\vthi}{v_{{\rm th}i}}

\newcommand{\vA}{v_\mathrm{A}}

\newcommand{\oA}{\omega_\mathrm{AW}}
\newcommand{\oS}{\omega_\mathrm{SW}}
\newcommand{\oK}{\omega_\mathrm{KAW}}
\newcommand{\oI}{\omega_\mathrm{ICW}}
\newcommand{\kparK}{k_{\parallel\mathrm{KAW}}}
\newcommand{\kparI}{k_{\parallel\mathrm{ICW}}}
\newcommand{\tcoll}{\tau_\nu}
\newcommand{\tnl}{\tau_\mathrm{nl}}

\newcommand{\vE}{\mbf{E}}
\newcommand{\vB}{\mbf{B}}

\newcommand{\vu}{\mbf{u}}
\newcommand{\vuperp}{\vu_\perp}

\newcommand{\vb}{\mbf{b}}

\newcommand{\vbperp}{\vb_\perp}

\newcommand{\vz}{\mbf{\hat z}}

\newcommand{\uperp}{u_\perp}

\newcommand{\jpar}{j_\parallel}

\newcommand{\upare}{u_{\parallel e}}
\newcommand{\bupare}{\overline{u}_{\parallel e}}
\newcommand{\tupare}{\widetilde{u}_{\parallel e}}
\newcommand{\upari}{u_{\parallel i}}
\newcommand{\bupari}{\overline{u}_{\parallel i}}
\newcommand{\tupari}{\widetilde{u}_{\parallel i}}

\title[Ion vs.\ electron heating by low-beta plasma turbulence]{Constraints on ion vs.\ electron heating by plasma turbulence at low beta}
\author[A.~A.~Schekochihin et al.] 
{A.~A.~Schekochihin,$^{1,2,3}$\thanks{Email: alex.schekochihin@physics.ox.ac.uk} 
Y.~Kawazura,$^{1}$\thanks{New address: Frontier Research Institute for Interdisciplinary 
Sciences and Department of Geophysics, Tohoku University, 
Aramaki aza Aoba 6-3, Aoba-ku, Sendai 980-8578, Japan.}
and M.~A.~Barnes$^{1,4,5}$\\
}

\affiliation{
$^1$Rudolf Peierls Centre for Theoretical Physics, University of Oxford, 
Clarendon Laboratory, Parks Road, Oxford OX1 3PU, UK\\[\affilskip]
$^2$Merton College, Oxford OX1 4JD, UK\\[\affilskip]
$^3$Niels Bohr International Academy, Blegdamsvej 17, 2100 Copenhagen, Denmark\\[\affilskip]
$^4$University College, Oxford OX1 4BH, UK\\[\affilskip]
$^5$United Kingdom Atomic Energy Authority, Culham Science Centre,
Abington~OX14~3DB,~UK
}

\begin{document}

\maketitle

\begin{abstract}
It is shown that in low-beta, weakly collisional 
plasmas, such as the solar corona, some instances of the solar wind,
the aurora, inner regions of accretion discs, their coronae, 
and some laboratory plasmas, Alfv\'enic fluctuations produce no ion heating within 
the gyrokinetic approximation, i.e., as long as their amplitudes (at the Larmor scale) 
are small and their frequencies stay below the ion Larmor frequency
(even as their spatial scales can be above or below the ion Larmor scale). 
Thus, all low-frequency ion heating in such plasmas is due to compressive 
fluctuations (``slow modes''): density perturbations and non-Maxwellian perturbations of the ion 
distribution function. 
Because these fluctuations energetically decouple from the Alfv\'enic ones 
already in the inertial range, the above conclusion means that the energy 
partition between ions and electrons in low-beta plasmas is decided at the 
outer scale, where turbulence is launched, and can be determined 
from magnetohydrodynamic (MHD) models of the relevant astrophysical systems. 
Any additional ion heating must come from non-gyrokinetic mechanisms such as cyclotron 
heating or the stochastic heating owing to distortions of ions' Larmor orbits. 
An exception to these conclusions occurs in the Hall limit, i.e., when the 
ratio of the ion to electron temperatures is as low as the ion beta 
(equivalently, the electron beta is order unity). 
In this regime, slow modes couple to Alfv\'enic ones 
well above the Larmor scale (viz., at the ion inertial or ion sound scale), so 
the Alfv\'enic and compressive cascades join and then separate again into two 
cascades of fluctuations that linearly resemble kinetic Alfv\'en and 
ion cyclotron waves, with the former heating electrons and the latter ions. 
The two cascades are shown to decouple, scalings for them are derived, and  
it is argued physically that the two species will be heated by them at approximately 
equal rates. 
\end{abstract}

\section{Introduction} 

One of the most fundamental questions in plasma astrophysics is what determines 
the temperatures of different particle species, ions ($T_i$) and electrons ($T_e$). 
We know that a system with different $T_i$ and $T_e$ is not in equilibrium 
and so must have an intrinsic (although not necessarily overwhelming) 
tendency to relax to an equi-temperature state. 
We do not, however, know of any 
mechanisms other than Coulomb collisions that would equalise the temperatures. 
There are no instabilities of a spatially homogeneous equilibrium 
with $T_i\neq T_e$ Maxwellian ions and electrons and so there is no obvious way in which, e.g.,
turbulence could result and quickly equalise the temperatures. In the absence of such 
fast dynamical processes, collisions are all that remains. In a large class of astrophysical 
and space plasmas where collisions are not very frequent, 
temperature equalisation by collisions is extremely slow: the relevant 
collision frequency is the ion-electron one, $\nu_{ie}$, which is a factor of mass ratio, 
$m_e/m_i$, smaller even than the electron collision frequency and a factor 
of $(m_e/m_i)^{1/2}$ smaller than the ion one. This means that, for most practical 
purposes, an ``incomplete'' equilibrium with $T_i\neq T_e$ must be 
assumed \citep[e.g.,][]{braginskii65}---and that is 
indeed what is observed in the solar wind \citep[see, e.g.,][and references therein]{cranmer09}. 
It is not, however, known what determines 
the ratio $T_i/T_e$---a question that is also of great interest in the context 
of extragalactic plasmas, e.g., accretion discs, 
where only $T_e$ is measured, but knowledge of $T_i$ is 
required for the understanding of basic plasma processes and model building
\citep[e.g.,][]{quataert03,sharma07,ressler17,rowan17,rowan19,chael18,chael19,chandran18}.  

Collisions aside, $T_i$ and $T_e$ can be changed via heating or cooling 
processes resulting from energy exchange between the mean (equilibrium) 
particle distributions and fluctuations (or waves), which are ubiquitously present 
in space and astrophysical plasmas---while temperature difference does not drive 
fluctuations, there are plenty of free-energy sources that do (background gradients, 
large-scale stirring, etc.). The free energy 
of these fluctuations is processed through phase space by various 
nonlinear (e.g., turbulence) and linear (e.g., phase mixing) mechanisms, 
brought to suitably small scales and thermalised,  
giving rise to ion or electron heating 
\citep[see, e.g.,][]{sch08,sch09,sch16,kawazura19,meyrand19}. 
The interesting question then is what fraction of the free energy injected 
at large scales is deposited into the thermal energy of each species. 

Much of the turbulence actually observed or theoretically expected 
in magnetised astrophysical plasmas is 
in the form of low-frequency, magnetohydrodynamic(MHD)-scale Alfv\'enic 
or compressive (``slow-mode'') fluctuations. They are at low frequencies because 
they are typically excited by large-scale mechanisms and because their cascade 
to smaller scales is anisotropic with respect to their local 
magnetic field, $\kpar\ll\kperp$, implying that the Larmor scales ($\rho_i$, $\rho_e$) in 
the perpendicular direction are reached before the Larmor frequencies ($\Omega_i$, $\Omega_e$) 
\citep[see][and references therein; this paper is henceforth referred to as S09]{sch09}. 
Thus, opportunities for transferring the energy of the turbulent cascade into 
ion thermal energy via the Landau damping of compressive fluctuations 
(throughout the inertial range; see S09--\S6 and \citealt{meyrand19})
or via the ion entropy cascade (starting at the ion Larmor scale; see S09--\S7 and \citealt{kawazura19}) 
occur before (i.e., at larger scales than) the cyclotron heating can take place \citep{howes08jgr}. 
All of these low-frequency heating routes can be treated in the so-called 
gyrokinetic (GK) approximation 
\citep[][the latter paper is henceforth referred to as H06]{frieman82,howes06}. 
This has the twin advantages 
of greater analytical tractability than the full Vlasov--Maxwell kinetics 
and very much greater feasibility of three-dimensional (3D) direct 
numerical simulations 
\citep{howes08prl,howes11prl,tenbarge13b,told15,banon16,li16,kawazura19}.\footnote{3D is 
the only relevant kind of simulations in this context because only in 3D can both the dominant 
nonlinearity and wave propagation be captured simultaneously \citep[see, e.g.,][]{howes15}. 
It is only in the last year 
that 3D full-Vlasov-kinetic simulations of the problem have become possible 
\citep{cerri18,groselj18prl,franci18,arzamasskiy19,zhdankin19}.} 
The goal of such analytical and numerical inquiries is to parametrise ion and electron heating 
in terms of the two main plasma parameters, $T_i/T_e$ and plasma beta, the latter defined 
to be the ratio 
of the ion thermal and magnetic energies, $\beta_i=8\pi n_iT_i/B^2$ (where $n_i$ is the ion 
density and $B$ is the magnetic field). 

Analytically, determining ion heating in anything like a definitive fashion 
has so far turned out to be a rather difficult task, 
except in the linear approximation \citep{quataert98,quataert99}. 
While progress can be made via modeling based on physically reasonable conjectures 
\citep[e.g.,][]{breech09,chandran09,chandran10,chandran10corona,howes10,howes11heating}, 
it is very useful to have some non-negotiable constraints on the answer, 
valid under clearly stated assumptions, such as the GK regime adopted here. 
In this paper, we show that it is possible to establish such constraints in a fairly straightforward 
way for low-beta plasmas, a subset that, while very far from being exhaustive, 
does include some observationally accessible cases, e.g., 
the solar corona \citep{aschwanden01,cranmer09rev}, 
some episodes of the solar wind at 1 AU \citep{smith01}, 
the aurora \citep{chaston08},
maybe, in the near future, certain regions of accretion discs \citep{chael18,chael19}, 
and, finally, laboratory plasmas such as 
the LAPD, custom-made for studies of Alfv\'en waves \citep{carter06,gekelman11}.  

In what follows, we will first, in \secref{sec:epitome}, give a qualitative physical outline 
of our argument and its implications. A reader uninterested in theoretical rigour 
need not read anything else. The subsequent three sections are dedicated 
to providing a systematic calculation to back up the statements made in \secref{sec:epitome}.
\Secref{sec:GK} is a quick recapitulation of the GK 
formalism needed in what follows. \Secref{sec:low_betae} contains 
the derivation of a reduced set of equations satisfied by 
plasma turbulence in the low-beta limit and the proof, based on those equations, 
that, in low-beta plasmas, there is no ion heating due to Alfv\'enic fluctuations
and that all ion heating that does occur is due 
to compressive fluctuations found in the inertial range. 
\Secref{sec:Hall} (whose length is perhaps incommensurate with 
its importance in the grand scheme of things) 
deals with a particular type of low-beta plasma where 
ions are much colder than electrons (the ``Hall limit''), where it turns out 
that the conclusion of \secref{sec:low_betae} must be substantially revised 
and the ion and electron heating rates are likely to be comparable
(on the way, some conceptually 
interesting results on Hall turbulence emerge: see \secref{sec:KAW_ICW}). 
Finally, in \secref{sec:disc}, there is a brief closing discussion, in particular of possible 
self-regulation mechanisms for ion heating. 
 
\section{Epitome}
\label{sec:epitome}

When turbulence in a plasma is stirred up by some large-scale mechanism, this amounts 
to ion and electron distribution functions being perturbed away from equilibrium. 
If these perturbations are low-frequency ($\omega\ll\Omega_i$) and 
large-scale ($k\rho_i\ll1$, where $\rho_i$ is the ion Larmor radius), 
they will, via nonlinear interactions, generate further, smaller-amplitude, 
smaller-scale, higher-frequency fluctuations. Just as in ordinary fluid 
turbulence, this process can be conceptualised as a cascade of energy---in 
the case of kinetic (collisionless or weakly collisional) plasma, 
a cascade of free energy associated with the perturbed distribution functions 
and electromagnetic fields (see \citealt{sch08} and references therein).  
In the presence of a strong magnetic field, the fluctuations produced 
by this cascade are expected---and, indeed, observed, in numerical simulations
\citep{cho00,maron01} and in the solar wind 
\citep{horbury08,podesta09aniso,wicks10,chenmallet11,chen16}---to 
be ever more scale-anisotropic at ever smaller scales, viz., long along the field, short 
across it:~$\kpar\ll\kperp$. 

In this anisotropic limit, the fluctuations at scales greater than $\rho_i$ 
can be classified into two kinds: 

(i) {\em Alfv\'enic}, i.e.,  
incompressible perpendicular MHD perturbations of the velocity and magnetic 
field, $\vuperp$ and $\vbperp$---these correspond to Maxwellian perturbations 
of the ion distribution function with flow velocity 
$\vuperp = c\vE\times\vB_0/B_0^2$, where $\vE$ is the perturbed electric field 
and $\vB_0$ the mean magnetic field (see S09--\S5);

(ii) {\em compressive}, i.e., 
perturbations of plasma density $\dn_e$ ($=\dn_i n_e/n_i$ by quasineutrality), 
field strength $\dB$, and general perturbations 
of the ion distribution function involving parallel flow velocity, temperature 
and higher moments (see S09--\S6); these perturbations are the kinetic 
version of the MHD slow modes.\footnote{High-frequency modes, such as 
fast magnetosonic waves (at inertial-range scales) or whistlers (at sub-ion-inertial scales), 
can be self-consistently 
ignored in the anisotropic regime that we are considering---and indeed are ordered 
out in the GK approximation (see H06--\S2.2). 
This does not mean that they cannot exist, just that they can not-exist, i.e., 
that they would not be triggered by the motions and fields that are retained.
Observations of solar-wind turbulence in the inertial range suggest that fast-wave energy 
is indeed negligible \citep{howes12}. Around the ion Larmor scale, an energetically 
subdominant population of non-GK 
perturbations, viz., whistler and ion-cyclotron waves with large $\kpar$, is 
observed in the solar wind \citep{wicks10,podesta11mhel,he11,he12,klein14,lion16} 
and may be due to pressure-anisotropy instabilities, which are not captured by GK, 
but are not a significant danger at low beta \citep[e.g.,][]{hellinger06,bale09,kunz18}.
\label{fn:nonGK}}

It is then possible to prove (S09--\S5, 6), 
for any $\beta_i$ and $T_i/T_e$ and assuming that the equilibrium distribution function 
is either Maxwellian or 
satisfies certain constraints \citep{kunz15}, that these two types of fluctuations 
are energetically decoupled from each other: the cascade of the free 
energy splits into an MHD ``Alfv\'en-wave cascade'' 
and a cascade of compressive fluctuations, passively advected by the 
Alfv\'en-wave turbulence but unable to exchange energy with it. 
In a weakly collisional plasma, i.e., one in which collision rates are small 
compared to the characteristic frequencies of the turbulent fluctuations, 
the latter cascade is potentially subject to Landau--Barnes damping \citep{Barnes66}, 
which will give rise to (parallel) ion heating at all scales in the 
``inertial range'' ($\kperp\rho_i<1$). However, 
the nonlinear cascade rate is comparable to the 
Alfv\'en frequency, $\kpar \vA$, where $\vA=B_0/\sqrt{4\pi m_in_i}$ is the 
Alfv\'en speed \citep{GS95,GS97}, 
whereas the damping rate cannot be much larger than $\sim\kpar\vthi$, 
where $\vthi=\sqrt{2T_i/m_i}$ is the ion thermal speed (S09--\S6.2.2).
In low-beta plasmas,  $\vthi=\sqrt{2T_i/m_i} = \sqrt{\beta_i} \vA \ll \vA$, 
so the damping is expected to be negligible compared to the rate of nonlinear 
transfer of the fluctuation energy towards the Larmor scale\footnote{Interestingly, 
it turns out that even at $\beta_i\sim1$, Landau damping in the inertial range 
can be effectively suppressed by a nonlinear effect, the stochastic echo \citep{meyrand19}, 
and the compressive free energy cascades mostly unimpeded towards the Larmor scale.} 
\citep[cf.][]{lithwick01}.

This means that (in a weakly collisional plasma) 
no thermalisation of any of that energy can occur until 
the fluctuations have reached $\kperp\rho_i\sim1$. At this point, 
the free-energy cascade is, in general, no longer split into 
Alfv\'enic and compressive, the two types of fluctuations can couple and 
the free energy can be shown to be cleanly split again into two decoupled 
cascades only at $\kperp\rho_i\gg1$. These two sub-Larmor dissipation channels 
are the cascades of kinetic Alfv\'en waves (KAW) and of ion entropy 
(S09--\S7; that the sub-Larmor-range turbulence in the solar wind is indeed 
predominantly of the KAW kind appears to be settled: 
see \citealt{salem12}, \citealt{chen13}, and the references in footnote~\ref{fn:nonGK}). 
The KAW eventually thermalise into electrons, via Landau damping 
and/or via various dissipation effects at or below the electron Larmor 
scale.\footnote{Because $\kpar\ll\kperp$, the frequency of the 
turbulent fluctuations at $\kperp\rho_i\sim1$ is still much smaller 
than $\Omega_i$. If the cyclotron frequency is reached by the KAW 
cascade at sub-Larmor scales, cyclotron heating can result, linearly, 
but the wave number range in which it occurs is quite narrow 
\citep[see Appendix of][]{howes08jgr} and it remains to be seen whether 
it would be effective at all in a nonlinear situation (see, however, 
\citealt{arzamasskiy19}). In any event, 
this heating mechanism is outside the scope of our treatment here; we 
acknowledge its possible contribution as a source of additional ion heating 
but remain agnostic about the amount of such heating.}  
The ion entropy cascade is a nonlinear mixing process in phase 
space, resulting in fine-scale structure in the ion distribution 
function and eventually thermalised into ions by collisions---as large 
gradients in $\vperp$ form alongside large spatial gradients, 
even very low collisionality is enough 
to dissipate non-Maxwellian perturbations at a finite rate (S09--\S\S7.9 and 7.10). 

Thus, how much energy goes into ions and how much into electrons 
is decided when the free-energy cascade reconstitutes itself and then 
splits again at the ion Larmor scale. Therefore, any analytical treatment of this 
problem requires a theoretical description uniformly 
valid for $\kperp\rho_i$ small, large and order-unity. Gyrokinetics 
is such a theory, requiring low frequencies ($\omega\ll\Omega_i$) but 
not long wavelengths (see H06 for a tutorial). However, 
in its general form, it does not make the problem of energy partition 
between species any more analytically tractable \citep[although it does make numerical 
simulations of this process more feasible:][the latter paper being the closest that we have got to an actual solution of the problem, at least within the GK approximation]{howes08prl,howes11prl,tenbarge13b,told15,banon16,li16,kawazura19}. 

A dramatic simplification occurs if $\beta_i$ is assumed small. 
In this limit, $\vthi/v_A=\sqrt{\beta_i}\ll1$, so ions cannot stream along the field lines 
fast enough to couple properly to the electromagnetic fields associated with the 
Alfv\'en waves, MHD or kinetic, and so perturbations of the ion 
distribution function (other than the Maxwellian $\vE\times\vB$ drift) 
stay decoupled from the Alfv\'enic cascade.\footnote{It is perhaps worth emphasising 
that it is the ion beta, $\beta_i$, that must be low, while $\beta_e$ may or may not be (the 
possibility that it is not is covered by the Hall limit; see the end of this 
section and \secref{sec:Hall}). The regime 
in which $\beta_i\sim1$ but $\beta_e\ll1$, i.e., electrons are colder than 
ions, $ZT_e/T_i\ll1$, is covered by the theory for order-unity or high $\beta_i$, 
which we do not attempt here (for a numerical study of what happens there, see 
\citealt{kawazura19}). 
The only difference between this regime and $\beta_e\sim\beta_i\sim1$ is that,
since $d_e = \rho_e/\sqrt{\beta_e}\gg\rho_e$, 
the electron inertial effects come in before the KAW cascade reaches the electron 
Larmor scale. This modifies the 
structure of the KAW cascade \citep{chen17,passot17}, but should not change 
the fact that all the energy that is processed through it goes into electrons. 
Such a physical situation is observable in the Earth's magnetosheath \citep{chen17}.
Interesting changes in the energy partition may be possible if $\beta_e$ is 
so low that $d_e\gtrsim \rho_i$, even though $\beta_i\sim1$. This is possible 
if $ZT_e/T_i \sim \beta_e \sim m_e/m_i$, perhaps too extreme a limit. We shall 
not consider it here. Note that the case of $\beta_e\sim m_e/m_i$ and 
$\beta_i\ll1$ (considered by \citealt{zocco11} and easy to simulate; 
see \citealt{loureiro13colless,loureiro16viriato,groselj17}) 
is no different, as far as energy partition is concerned,  
from the standard low-beta regime---Alfv\'enic fluctuations heat electrons, 
compressive ones heat ions (see \secref{sec:ultra_low}).} 
As we shall demonstrate below, this makes it possible to prove that, 
in the low-beta limit, compressive perturbations of the 
ion distribution function will cascade from the inertial range, 
through the ion Larmor scale and turn into the ion entropy 
cascade at sub-Larmor scales and then into ion heat
without exchanging any energy with the Alfv\'enic fluctuations. 
All of the energy of the latter turns into KAW energy, which includes 
density perturbations at sub-Larmor scales, but cascades 
separately from the ion entropy and is eventually dissipated on electrons. 
Proving this analytically is 
accomplished by showing that a certain form of the free-energy invariant, which reduces 
to the energy of Alfv\'enic and KAW perturbations in the long- and short-wavelength 
limits, respectively, is conserved across the ion-Larmor-scale transition and thus no 
Alfv\'enic energy can leak into ion heat (see \secref{sec:low_betae}). 
Therefore, only the energy of what started out as compressive cascade 
in the inertial range will contribute to ion heating, at least to the extent  
that the GK approximation holds. 
Any further ion heating will have to come from non-GK mechanisms such 
as cyclotron heating \citep{gary05,kasper08,kasper13,marsch11,arzamasskiy19} 
or stochastic orbit deformations 
\citep[][see \secref{sec:sh}]{chandran10,chandran10corona,vech17,mallet18,arzamasskiy19}. 

The key ``practical'' (in astrophysics, this means relevant to large-scale modelling) 
conclusion from all this is that {\em at low $\beta_i$, the energy partition 
between ions and electrons is determined already at the outer scale of the 
MHD cascade}, where the energy flux splits into Alfv\'enic and compressive. 
Once this separation occurs, the ratio between ion and electron heating rates is fixed. 
Thus, what in principle is a microscale kinetic effect is in fact fully 
constrained by fluid dynamics.\footnote{This is, of course, only true assuming that 
the outer scale is collisional, 
so the Alfv\'enic and compressive cascades split within the MHD approximation 
and the transition to collisionless regime occurs within the compressive 
cascade (see S09--\S6). If the plasma is collisionless already at the outer 
scale, how the cascades separate is a fully kinetic problem, even if still 
a large-scale~one. Furthermore, in such a plasma, if $\dB/B \sim1$ at the outer 
scale and beta is low, the fluctuation energy far exceeds the thermal energy 
and one can hardly assume that a two-temperature Maxwellian equilibrium  
would either be established or survive. We are not addressing here the ``violent 
relaxation'' of such situations.} 
Since all the action is at the outer scale, the ion-to-electron heating ratio 
may depend on various nonuniversal circumstances, e.g., presence of equilibrium 
temperature stratification (which will produce temperature perturbations), 
shear, rotation, configuration of magnetic field, etc. 

These conclusions hold provided $\beta_e\sim\beta_i\ll1$. 
When $\beta_i\ll1$ but $\beta_e\sim 1$, i.e., when ions are much colder than electrons, 
$ZT_e/T_i\gg1$ (the so-called Hall limit; S09--\S E), 
the situation changes substantially (\secref{sec:Hall}). 
The physical difference between the $ZT_e/T_i\sim1$ and $ZT_e/T_i\gg1$ cases 
is that in the latter limit, slow magnetoacoustic waves are faster than 
the ions (because both the Alfv\'en speed 
and the sound speed $\cs = \sqrt{ZT_e/m_i}$ are larger than $\vthi$), 
remain undamped, and join happily with the AW cascade 
at a certain transition scale that is larger than $\rho_i$ 
[it is either the ion inertial or ion sound scale; see \exref{eq:rhoh_def}]. 
At this transition scale, the Alfv\'enic and compressive cascades 
re-couple and, below the transition scale, turn into cascades of higher-frequency KAW 
(or, as they are sometimes called in the context of Hall MHD, whistlers) 
and lower-frequency oblique ion cyclotron waves (ICW). 
In \secref{sec:KAW_ICW}, we argue, with some support from numerical 
simulations \citep{meyrand18}, that the two cascades are 
energetically decoupled and critically balanced, enabling one to predict 
their scalings easily: the KAW scalings [see \exref{eq:spectra_KAW}]
are the usual ones, as derived in S09--\S7.5 and \citet{cho04}; 
the ICW scalings [see \exref{eq:spectra_ICW}] are the same as 
the scalings for the inertial-wave turbulence proposed by \citet{nazarenko11} 
(both sets of of scalings can be related to some of the spectra previously 
posited by \citealt{krishan04} and by \citealt{galtier07}, and 
seen numerically by \citealt{meyrand12}).     
It is then possible to prove rigorously that the KAW cascade will heat electrons 
and the ICW cascade will heat ions---because the latter turns into 
the ion-entropy cascade (S09--\S7.10), whereas the former transitions through the 
ion Larmor scale without coupling to ions and turns into the sub-Larmor KAW 
cascade (this is all worked out in detail and further discussed in \secref{sec:Hall_Larmor}). 
We also argue, on physical grounds rather than rigorously, 
that the partition of energy flowing into the two cascades and, therefore, 
into the two species, should be approximately equal, independently of 
how much of the original MHD cascade is Alfv\'enic and how much compressive 
(\secref{sec:Hall_heating}). 
Here again, the energy partition is decided at fluid scales, but 
at the Hall transition scale rather than at the outer scale.  

The following sections are mostly technical in nature (except for 
\secsref{sec:KAW_scalings}, \ref{sec:ICW_scalings} and \ref{sec:Hall_heating}) 
and dedicated to proving the statements made above. 
In the process, we also derive a number of simple and 
appealing reduced models of various types of kinetic turbulence, which can be studied 
analytically and numerically using these models, either for its own sake or 
with some applied purpose. A reader uninterested in analytical detail 
can now skip to \secref{sec:disc}. 

\section{Gyrokinetic Primer} 
\label{sec:GK}

This section is an extended recapitulation of the GK formalism 
that is required for subsequent developments. In principle, all of this 
is already available from H06 and S09 (and, in a form generalised to non-Maxwellian 
equilibria, from \citealt{kunz15,kunz18}), but we provide this refreshed version 
for the convenience of the reader and as an opportunity to adjust notation, 
to deal with some subtleties, 
and to cast some of the derivations in what we now consider a more optimal form. 
A reader familiar with H06 and S09 may wish to skip (or skim) this section and then refer 
back to it as required during the reading of the rest of the paper. 

\subsection{Notation: Alfv\'enic Fields}

The electric and magnetic fields are described by the scalar potential $\phi$ 
and vector potential $\vAA$. It is convenient to introduce dimensionless versions 
of $\phi$ and of the component of $\vAA$ parallel to the equilibrium magnetic 
field $\vB_0 = B_0\vz$: 
\beq
\label{eq:phi_A_def}
\ephi = \frac{Ze\phi}{T_i} = \frac{2\Phi}{\rho_i\vthi},
\quad
\A = \frac{\Apar}{\rho_i B_0} = - \frac{\Psi}{\rho_i\vA},
\eeq
where $-e$ is the electron charge, $Ze$ the ion charge, $T_i$ the ion 
equilibrium temperature, $\vthi = \sqrt{2T_i/m_i}$ the ion thermal speed, 
$m_i$ the ion mass, $\rho_i=\vthi/\Omega_i$ the ion Larmor radius, 
$\Omega_i = ZeB_0/m_i c$ the ion Larmor frequency, $c$ the speed of light, 
$\vA = B_0/\sqrt{4\pi m_i n_i}$ the Alfv\'en speed, and
$n_i$ the equilibrium ion density. In what follows, we shall drop the ion 
species index everywhere except for some iconic quantities (e.g., $\rho_i$) 
or where there is a possibility of ambiguity (e.g., $T_i$ vs.\ $T_e$). 

In the above, we have also introduced the stream function $\Phi$ ($=c\phi/B_0$) 
of the $\vE\times\vB$ flow associated with $\phi$ and the flux function 
$\Psi$ giving (in velocity units) the magnetic-field perturbation perpendicular to $\vB_0$: 
\beq
\vuperp = \vz\times\vdperp\Phi,\quad
\vbperp = \vz\times\vdperp\Psi. 
\label{eq:PhiPsi_def}
\eeq
Physically these perturbations are 
Alfv\'en waves (AW). In the inertial range of magnetised plasma turbulence, 
they decouple from all other modes (the fast modes, which 
are ordered out in the GK approximation, and the slow, or compressive, modes) 
and satisfy the ``Reduced MHD'' 
equations (RMHD, first derived by \citealt{kadomtsev74} and \citealt{strauss76}; 
for a GK derivation, see S09--\S5.3):
\beq
\frac{\dd\Psi}{\dd t} = \vA\dpar\Phi,\quad
\frac{\rmd}{\rmd t}\dperp^2\Phi = \vA\dpar\dperp^2\Psi.
\label{eq:RMHD}
\eeq
Here the nonlinearities are hidden in the convective time derivative 
and in the spatial derivative along the perturbed field lines: 
\begin{align}
\label{eq:dt_def}
\frac{\rmd}{\rmd t} &= \frac{\dd}{\dd t} + \vuperp\cdot\vdperp 
= \frac{\dd}{\dd t} + \{\Phi,\dots\} 
= \frac{\dd}{\dd t} + \frac{\rho_i\vthi}{2}\{\ephi,\dots\},\\
\label{eq:dpar_def}
\dpar &= \frac{\dd}{\dd z} + \frac{\vbperp}{\vA}\cdot\vdperp 
= \frac{\dd}{\dd z} + \frac{1}{\vA}\{\Psi,\dots\}
= \frac{\dd}{\dd z} - \rho_i\{\A,\dots\},
\end{align}
where $\{f,g\}=(\dd_x f)(\dd_y g) - (\dd_x g)(\dd_y f)$. These derivatives 
will appear ubiquitously in what follows. 

\subsection{Gyrokinetic Equation}

Our starting point is standard, slab, Maxwellian gyrokinetics (see the derivation in H06 
or a summary in S09--\S3). In it, the ion distribution function is represented as 
\beq
f = F_0 + \df,\quad \df = -\ephi(\vr) F_0 + h(\vR), \quad \vR = \vr + \vrho,\quad 
\vrho = \frac{\vvperp\times\vz}{\Omega}, 
\label{eq:fi}
\eeq
where $\vR$ is the GK spatial coordinate (centre of Larmor ring), whereas $\vr$ 
is the usual spatial coordinate (position of the particle). 
Then the GK equation for the evolution of $h$~is
\beq
\frac{\dd h}{\dd t} + \vpar\frac{\dd h}{\dd z} + 
\frac{\rho_i\vth}{2}\lt\{\la\chi\ra_\vR,h\rt\}
= \frac{\dd\la\chi\ra_\vR}{\dd t} F_0 + C[h]
+ \frac{2\vpar\la\aext\ra_\vR}{\vth^2} F_0. 
\label{eq:h}
\eeq
Here the GK potential gyroaveraged at constant $\vR$ (an operation denoted by angle brackets)~is 
\beq
\la\chi\ra_\vR = \frac{Ze}{T_i}\lt\la \phi - \frac{\vv\cdot\vAA}{c}\rt\ra_\vR 
= \hJ_0\ephi - 2\hvpar\hJ_0\A + \hvperp^2\hJ_1\frac{\dB}{B}, 
\eeq 
where $\hvpar = \vpar/\vth$, $\hvperp=\vperp/\vth$, 
$\dB$ is the perturbation of the magnetic field along itself (related to $\vAperp$), 
which is also the perturbation of the field's strength; 
the gyroaveraging Bessel operators 
are defined in terms of their Fourier space representations:
\beq
\hJ_0 \leftrightarrow J_0(a) = 1 - \frac{a^2}{4} + \dots, \quad 
\hJ_1 \leftrightarrow \frac{2J_1(a)}{a} = 1 - \frac{a^2}{8} + \dots,\quad
a = \frac{\kperp\vperp}{\Omega} = \hvperp\kperp\rho_i. 
\eeq
Obviously, $a^2\leftrightarrow -\hvperp^2\rho_i^2\dperp^2 = -\hvperp^2\hdperp^2$,
where we denote $\hdperp=\rho_i\dperp$. 
We shall use the $\hat J$ notation \citep{kunz18} interchangeably with $\la\dots\ra_\vR$ 
(or with $\la\dots\ra_\vr$, the gyroaverage of an $\vR$-dependent 
quantity at constant $\vr$), as proves convenient. 

The last term in \exref{eq:h} represents energy injection by means of an external 
parallel acceleration $\aext$. This will be a convenient model of the excitation 
of compressive perturbations for further calculations dealing with free-energy budgets.  
Finally, the collision operator $C[h]$ contains both the ion-ion and ion-electron 
collisions, but the latter are negligible in the mass-ratio expansion adopted below. 

\subsection{Isothermal Electron Fluid}
\label{sec:IEF}

We supplement the ion GK equation \exref{eq:h} with two fluid equations 
arising from the isothermal approximation for electrons, which is a result of expansion 
in the electron-ion mass ratio and holds at $\kperp\rho_e\ll1$,\footnote{This is as good 
a place as any to address a certain 
resentment that a reader with a predilection for mathematical rigour \citep[e.g.,][]{eyink15,eyink18}
might experience towards approximate equations valid in restricted scale subranges. 
Generally speaking, nonlinear solutions 
of such approximate equations will not stay within their own bounds of validity 
and develop gradients on scales that are smaller than allowed by the assumed ordering. 
This is, of course, what turbulence does, or, indeed, is: a cascade to smaller scales, 
in pursuit of dissipation. In such a cascade, the smallest scales are typically 
reached in $\sim$ one turnover time, 
regardless of how wide the full range of available scales is. Therefore, formally, any 
system of equations restricted to a subrange of scales is only valid for $\sim$ one turnover 
time; ``non-ideal'' effects associated with dissipation at smaller scales 
come in after that (e.g., ideal-MHD solutions do not stay ideal 
for long, however small is the resistivity or other flux-unfreezing effects; 
see, e.g., \citealt{boozer18}). 
This limited validity is, however, sufficient for analysing basic linear and nonlinear 
interactions that govern the transfer of energy through the scale subrange that is 
under consideration, as long as this transfer can be assumed local to this subrange. 
The approximate equations can also be usefully simulated numerically as long as 
some regularisation at small scales is provided and assuming that the nature of this 
regularisation is unimportant---i.e., that as long as a free-energy sink is present 
at the smallest scale of the considered subrange, 
its detailed microphysics does not affect the behaviour of larger scales 
(this, of course, does not always have to be the case, but tends to be).}  
and with field equations that follow from quasineutrality and Amp\`ere's law in the same 
approximation (this system of equations was derived in S09--\S4, implemented 
numerically by \citealt{kawazura18} and simulated to some useful effect 
by \citealt{kawazura19}):
\begin{align}
\label{eq:A}
&\frac{\dd\A}{\dd t} + \frac{\vth}{2}\dpar\ephi 
= \frac{\vth}{2}\dpar\frac{Z}{\tau}\frac{\dn}{n} + \eta\dperp^2\A,\\ 
\label{eq:cont}
&\frac{\rmd}{\rmd t}\lt(\frac{\dn}{n} - \frac{\dB}{B}\rt) + \dpar\upare 
= -\frac{\rho_i\vth}{2}\lt\{\frac{Z}{\tau}\frac{\dn}{n},\frac{\dB}{B}\rt\},\\
\label{eq:quasineut}
&\frac{\dn}{n} = -\ephi + \overline{\hJ_0 h},\\
\label{eq:Amppar}
&\frac{\upare}{\vth} = \frac{1}{\beta_i}\hdperp^2\A + \overline{\hvpar\hJ_0 h} + \Jext,\\
\label{eq:Ampperp}
&\frac{2}{\beta_i}\frac{\dB}{B} = \lt(1+\frac{Z}{\tau}\rt)\ephi 
- \frac{Z}{\tau}\overline{\hJ_0 h} - \overline{\hvperp^2\hJ_1 h}.
\end{align} 
Here $\tau=T_i/T_e$, $\dn/n$ is the relative electron density perturbation (which is the same 
as the ion one, by quasineutrality), $\upare$ the parallel electron flow velocity, 
and overlines denote velocity integrals: 
$\overline{(\dots)} = (1/n_i)\int\rmd^3\vv(\dots)$;  
note that the integrals are taken at constant $\vr$ and so $\vR$-dependent 
quantities under them must be gyroveraged at constant $\vr$, hence the appearance 
of the $\hJ_0$ and $\hJ_1$ operators. In the above system, \exref{eq:A} 
is the parallel component of Ohm's law (electron's momentum equation), 
\exref{eq:cont} is the electron continuity equation, 
\exref{eq:quasineut} is the statement of quasineutrality, 
\exref{eq:Amppar} and \exref{eq:Ampperp} are the parallel and 
perpendicular components, respectively, of Amp\`ere's law (the perpendicular 
one is equivalent to the statement of perpendicular pressure balance; 
see S09--\S3.3). 

In the right-hand side of \exref{eq:Amppar}, we have added a 
forcing term for the Alfv\'enic perturbations, which can be viewed 
as arising from a model external (nondimensionalised) ``AW-antenna'' 
current $\Jext \equiv \jext/en_e\vth$ \citep[e.g.,][]{tenbarge14}. In the 
right-hand side of \exref{eq:A}, 
we have added a resistive term ($\eta$ is the magnetic diffusivity) 
to represent dissipation of energy into electron heat and to allow flux unfreezing 
at small scales \citep[an important concern: see][]{eyink15,eyink18,boozer18}. 
Formally, this effect is outside the mass-ratio 
ordering that gave us the hybrid equations introduced above and 
would have to be brought in alongside electron inertia and electron-collisional effects 
(see S09--\S7.12 and \citealt{zocco11}), but we can treat the resistive 
term as a representative for all of that as far as magnetic reconnection 
and free-energy thermalisation on electrons are concerned.\footnote{\Eqref{eq:A} is the electron 
parallel momentum equation, representing the force balance between the parallel electric 
field (the left-hand side), parallel pressure gradient (the first term on the right-hand side) 
and the collisional drag force, which is the resistive term. Technically speaking, the latter 
is proportional to the difference between the electron velocity $\upare$ and the ion 
velocity $\upari=\overline{\vpar\hJ_0 h}$, which is worked out from \exref{eq:Amppar}. 
Including normalisations, the resistive term is then
$\nu_{ei}(\upare-\upari)cm_e/e\rho_i B_0 = \eta\lt(\dperp^2\A + \beta_i\Jext/\rho_i^2\rt)$, 
where $\nu_{ei}$ is the electron-ion collision frequency and $\eta = \nu_{ei} d_e^2$.
However, if $\eta$ is small, it will only matter when it multiplies $\dperp^2$, as 
$\A$ develops small-scale structure. Since we assume $\Jext$ to be a 
large-scale quantity, it can be dropped wherever it multiplies~$\eta$.\label{fn:eta_jext}} 
This is reasonable because the precise details of how the energy is removed from 
the system should not matter, so long as it happens at scales smaller than 
the ion Larmor scale and does not introduce artificial coupling between ions 
and electrons.\footnote{A minor nuance is that, in numerical practice, resistivity 
alone is usually insufficient to terminate a turbulent cascade described by 
\exref{eq:A} and \exref{eq:cont}---one must have a small-scale regularisation 
term in \exref{eq:cont} as well \citep{kawazura18}. Formally, such a term 
would represent collisionless and/or collisional electron damping at and below 
the electron Larmor scale. This too is electron heating. For the purposes of 
analytical energy budgets considered in this paper, the resistive term is a sufficient 
representative for~it.} 
These features---forcing and resistivity---will be useful in working out free-energy budgets.

\subsection{Free-Energy Budget}
\label{sec:free_energy}

The $\df$ gyrokinetics conserves (except for explicit sources and sinks) 
a quadratic norm of the perturbations, known as the free energy (see S09--\S3.4 
and references therein): 
\beq
W = \int\frac{\rmd^3\vr}{V}\lt(\sum_s\int\rmd^3\vv\frac{T_s\df_s^2}{2 F_{0s}} 
+ \frac{|\dvB|^2}{8\pi}\rt),
\eeq  
where $V$ is the volume of the plasma. 
Here the perturbed ion distribution function is given by \exref{eq:fi}, 
the perturbed electron distribution function under the mass-ratio expansion 
is $\df_e = (\dn/n) F_{0e}$ (see S09--\S4.4), and so, in our notation, 
\beq
W = \frac{\vth^2}{4}\int\frac{\rmd^3\vr}{V}\lt[\frac{\overline{\la h^2\ra_{\vr}}}{F_0} 
-\ephi^2 - 2\ephi\frac{\dn}{n} + \frac{Z}{\tau}\frac{\dn^2}{n^2} 
+ \frac{2}{\beta_i}\lt(|\hvdperp\A|^2 + \frac{\dB^2}{B^2}\rt)\rt],
\label{eq:W_h}
\eeq 
where we have dropped the prefactor of $m_i n_i$. 

Since $\int\rmd^3\vr\overline{\la h^2\ra_{\vr}/F_0} = \overline{\int\rmd^3\vR\, h^2/F_0}$, 
we may derive the evolution equation for $W$ by multiplying \exref{eq:h} by $h/F_0$, 
integrating over the entire GK phase space and using \exsdash{eq:A}{eq:Ampperp} 
opportunely. The result is 
\beq
\frac{\rmd W}{\rmd t} = \epsA + \epsC - Q_i - Q_e,
\label{eq:W_evo}
\eeq
where the sources are the injection rates of the Alfv\'enic and compressive perturbations, 
\begin{align}
\label{eq:epsA_def}
\epsA &= \vth^2 \int\frac{\rmd^3\vr}{V}\frac{\dd\A}{\dd t} \Jext
= -\int\frac{\rmd^3\vr}{V}\,\vE\cdot\vjext,\\ 
\epsC &= \int\frac{\rmd^3\vr}{V}\,\aext\overline{\vpar\hJ_0 h} = \int\frac{\rmd^3\vr}{V}\,\aext\upari,
\label{eq:epsC_def}
\end{align}
and the sinks are the ion and electron heating rates:\footnote{If we had retained 
the $\eta\Jext$ term in \exref{eq:A} (dismissed in footnote~\ref{fn:eta_jext}) 
and the ion-electron part of $C[h]$, the electron heating term would have turned out 
to be  $Q_e = \nu_{ei}(Zm_e/m_i)\int\rmd^3\vr (\upare-\upari)^2/V = 
(4\pi\eta/c^2) \int\rmd^3\vr\jpar^2/V$, the total Ohmic heating. 
This is the same as \exref{eq:Qe_def} if we drop all terms that 
are small in the mass-ratio expansion, only retaining instances of $\eta$ 
multiplying the highest spatial derivatives of $\A$.} 
\begin{align}
\label{eq:Qi_def}
Q_i &= -\frac{\vth^2}{2}\overline{\int\frac{\rmd^3\vR}{V} \frac{hC[h]}{F_0}}\ge0,\\
Q_e &= \eta\frac{\vth^2}{\beta_i\rho_i^2} \int\frac{\rmd^3\vr}{V}\,|\hdperp^2\A|^2 
= \eta \int\frac{\rmd^3\vr}{V} \frac{|\dperp^2\Apar|^2}{4\pi m_i n_i}\ge0. 
\label{eq:Qe_def}
\end{align} 
We have restored dimensions these expressions to make their physical meaning 
more transparent. Note that in \exref{eq:epsA_def}, the final expression---the work done 
by the electric field against the external current---is obtained by noticing that 
there is a perpendicular current associated with $\jext$, which is small in the GK 
expansion (because, to avoid injecting charges, $\vdel\cdot\vjext = 0$) and so, to lowest 
order in $\kpar/\kperp$, 
\beq
\int\frac{\rmd^3\vr}{V}\frac{1}{c}\frac{\dd\Apar}{\dd t}\jext =  
\int\frac{\rmd^3\vr}{V}\frac{1}{c}\frac{\dd\mbf{A}}{\dd t}\cdot\vjext = 
\int\frac{\rmd^3\vr}{V}\lt(\frac{1}{c}\frac{\dd\mbf{A}}{\dd t} + \vdel\phi\rt)\cdot\vjext, 
\eeq   
with the expression for $-\vE$ able to be completed with $\vdel\phi$ under the integral 
because $\vdel\cdot\vjext = 0$.

In steady state, \exref{eq:W_evo} is the overall free-energy budget, 
which says that the total injection is equal to the total dissipation. 
The main purpose of this paper is to work out more restrictive 
energy budgets that constrain $Q_i$ and $Q_e$ separately. 

\subsection{Separating Alfv\'enic Perturbations}
\label{sec:IEF_g}

We now rearrange the perturbed distribution function in a way that 
has the effect of separating the Alfv\'enic part 
of the distribution function from its ``compressive'' part:\footnote{Note that 
this is a different rearrangement than in S09--\S5.1 and so the subsequent derivation, 
while similar in spirit to S09--\S5, is different in detail. We shall see that 
this is a more convenient approach.} 
\beq
\label{eq:g_ansatz}
h = \la\ephi\ra_\vR F_0 + g = \hJ_0\ephi F_0 + g
\hence
\df = (\la\ephi\ra_\vR - \ephi)F_0 + g,\quad
g = \la\df\ra_\vR. 
\eeq
The field equations \exsdash{eq:quasineut}{eq:Ampperp} become 
\begin{align}
\label{eq:dn}
&\frac{\dn}{n} = -(1-\hG_0)\ephi + \overline{\hJ_0 g},\\
\label{eq:upar}
&\frac{\upare}{\vth} = \frac{1}{\beta_i}\hdperp^2\A + \overline{\hvpar\hJ_0 g} + \Jext,\\
\label{eq:dB}
&\frac{2}{\beta_i}\frac{\dB}{B} = - \frac{Z}{\tau}\frac{\dn}{n} + (1-\hG_1)\ephi 
- \overline{\hvperp^2\hJ_1 g},
\end{align}
where two more Bessel operators have arisen: 
\begin{align}
\label{eq:G0}
\hG_0 & \leftrightarrow \overline{J_0^2(a) F_0} = I_0(\alpha)e^{-\alpha} 
= 1 - \alpha + \dots,
\quad \alpha = \frac{\kperp^2\rho_i^2}{2} \leftrightarrow - \frac{1}{2}\hdperp^2,\\
\hG_1 & \leftrightarrow \overline{\hvperp^2\frac{2J_1(a)J_0(a)}{a}F_0} 
= -\lt[I_0(\alpha)e^{-\alpha}\rt]' = 1 - \frac{3}{2}\alpha + \dots\ .
\label{eq:G1}
\end{align}
The GK equation \exref{eq:h}, rewritten in terms of $g$, becomes 
\begin{align}
\nonumber
&\frac{\dd}{\dd t}\lt(g - \hvperp^2\hJ_1\frac{\dB}{B}F_0\rt) 
+ \frac{\rho_i\vth}{2}
\lt(\lt\{\la\ephi\ra_\vR, g - \hvperp^2\hJ_1\frac{\dB}{B}F_0\rt\}
+ \lt\{\hvperp^2\hJ_1\frac{\dB}{B}, g\rt\}\rt)\\
\nonumber
&\qquad+ \vpar\lt\la\dpar\lt(g + \frac{Z}{\tau}\frac{\dn}{n} F_0\rt)
+ \rho_i\{\A - \la\A\ra_\vR,\ephi-\la\ephi\ra_\vR\}F_0\rt\ra_\vR  \\
&\qquad\qquad= C\lt[g + \la\ephi\ra_\vR F_0\rt] + \frac{2\vpar\la\aext\ra_{\vR}}{\vth^2}F_0. 
\label{eq:g}
\end{align}
This has been derived by using \exref{eq:A} to express $\la\dd\A/\dd t\ra_\vR$
and after some manipulation of gyroaverages.\footnote{If we 
are to be consistent, we must retain in \exref{eq:g} a forcing term associated with the 
resistive term in \exref{eq:A}. As we explained in footnote~\ref{fn:eta_jext}, the 
full form of this resistive term is 
$\nu_{ei}(\upare-\upari)cm_e/e\rho_iB_0 = \nu_{ie}(\upare-\upari)/\vth$, 
where $\nu_{ie} = (m_en_e/m_in_i)\nu_{ei}$ is the ion-electron collision 
frequency. The additional term that belongs in the left-hand side of \exref{eq:g} 
is, therefore, $2\nu_{ie}\vpar\la\upare-\upari\ra_{\vR}F_0/\vth^2$, which is  
minus the ion-electron friction force. But this is cancelled by the linearised 
ion-electron collision operator, which, to lowest order in the mass-ratio 
expansion, is just the ion-electron friction (see, e.g., \citealt{helander05}). 
Thus, from now on, we may drop the resistive term in \exref{eq:g} as long as 
the collision operator in this equation is understood to contain the ion-ion 
collisions only.}  

In terms of $g$, the free energy \exref{eq:W_h} becomes 
\beq
W = \frac{\vth^2}{4}\int\frac{\rmd^3\vr}{V}\lt[\frac{\overline{\la g^2\ra_{\vr}}}{F_0} 
+ \ephi(1-\hG_0)\ephi + \frac{Z}{\tau}\lt|(1-\hG_0)\ephi - \overline{\hJ_0 g}\rt|^2
+ \frac{2}{\beta_i}\lt(|\hvdperp\A|^2 + \frac{\dB^2}{B^2}\rt)\rt], 
\label{eq:W_g}
\eeq
where, by definition, 
$(1/V)\int\rmd^3\vr\ephi(1-\hG_0)\ephi = \sum_\vk(1-\Gamma_0)|\ephi_\vk|^2$. 

For some upcoming derivations, it will be useful to have the zeroth moment 
of \exref{eq:g}. We integrate \exref{eq:g} over 
velocities at constant $\vr$, use \exref{eq:dn} to express $\overline{\hJ_0 g}$ 
and subtract \exref{eq:cont} from the resulting equation, 
using \exref{eq:upar} for $\upare$ and so far neglecting nothing. 
The outcome~is
\begin{align}
\nonumber
&\frac{\rmd}{\rmd t}\lt[(1-\hG_0)\ephi + (1-\hG_1)\frac{\dB}{B}\rt] 
- \vth\dpar\lt(\frac{1}{\beta_i}\hdperp^2\A + \Jext\rt)= 
\rho_i\overline{\lt\la\{\la\A\ra_\vR-\A,\vpar g\}\rt\ra_\vr}\\
\nonumber
&\qquad - \frac{\rho_i\vth}{2}\lt[
\overline{\lt\la\lt\{\la\ephi\ra_\vR - \ephi, g - \hvperp^2\hJ_1\frac{\dB}{B}F_0\rt\}
+ \lt\{\hvperp^2\hJ_1\frac{\dB}{B}, g\rt\}\rt\ra_\vr}
- \lt\{\frac{Z}{\tau}\frac{\dn}{n},\frac{\dB}{B}\rt\}\rt]\\
&\qquad + \overline{\la C[g+\la\ephi\ra_\vR F_0]\ra_\vr}.
\label{eq:gint}
\end{align}
Note that \exref{eq:A} and \exref{eq:gint} have the makings of the RMHD system \exref{eq:RMHD}: 
this emerges from any long-wavelength 
approximation where one can neglect the $\dn$ term in \exref{eq:A}, 
as well as $(1-\hG_1)\dB/B$ and all of the right-hand side of \exref{eq:gint}. 
This is indeed how RMHD is derived from gyrokinetics in the limit of 
$\kperp\rho_i\ll1$ (S09--\S\S5.2, 5.3). Below, we shall apply somewhat different orderings 
to work out the reduced dynamics at low beta. 

\section{Reduced Dynamics and Heating at Low Beta} 
\label{sec:low_betae}

We shall now show that no ion heating occurs in the low-beta regime, viz., at $\beta_i\ll1$. 
The problem has two governing parameters, $\beta_i$ and $\beta_e=Z\beta_i/\tau$. 
There are two interesting limits:
 
(i) $\beta_e\sim\beta_i\ll1$ ($Z/\tau\sim1$)---this section,

(ii) $\beta_e\sim1$ and $\beta_i\ll1$ ($Z/\tau\sim\beta_i^{-1}\gg1$, cold ions)---\secref{sec:Hall}.
 

\noindent In considering these limits, we shall make use of the notation and equations 
introduced in \secref{sec:GK}. Namely, our starting point is the system 
consisting of the six equations 
\exsdash{eq:A}{eq:cont}, \exsdash{eq:dn}{eq:dB} and \exref{eq:g} 
for six fields $\A$, $\ephi$, $\dn$, $\dB$, $\upare$ and $g$. 

\subsection{Ordering}
\label{sec:low_beta_ordering}

Working in the limit $\beta_e\sim\beta_i\ll1$,  we let 
\beq
\frac{Z}{\tau} = \frac{\beta_e}{\beta_i} \sim 1, \quad \kperp\rho_i \sim 1, 
\label{eq:krho_one}
\eeq
the latter assumption meaning that we are able to treat the Larmor-scale transition directly. 

Since we wish to be able to handle Alfv\'enic perturbations, and since 
we wish their linear frequency ($\kpar\vA$) and their nonlinear 
interaction rate ($\kperp\uperp$) to be able to be comparable, 
we stipulate  
\beq
\frac{\dBperp}{B} \sim \frac{\uperp}{\vA} \sim \frac{\kpar}{\kperp} \sim \epsilon,
\label{eq:GKbasic}
\eeq
where $\epsilon$ is the basic GK expansion parameter, 
with no further $\beta_i$-related factors, of which we shall now keep a close watch. 
In view of \exref{eq:krho_one}, this assumption implies 
\begin{align}
\label{eq:A_order}
\frac{\dBperp}{B} \sim \frac{\kperp\Apar}{B_0} \sim \kperp\rho_i\A 
&\hence \A \sim \epsilon,\\
\frac{\uperp}{\vA} \sim \frac{c\kperp\phi}{\vA B_0} \sim \kperp\rho_i\sqrt{\beta_i}\,\ephi
&\hence \ephi \sim \frac{\epsilon}{\sqrt{\beta_i}}. 
\label{eq:phi_order}
\end{align}
Examination of \exsdash{eq:dn}{eq:dB} then suggests that 
\beq
\frac{\dn}{n}\sim\frac{g}{F_0}\sim\ephi\sim\frac{\epsilon}{\sqrt{\beta_i}},\quad
\frac{\dB}{B}\sim\epsilon\sqrt{\beta_i}. 
\label{eq:low_beta_order}
\eeq 

\subsection{Equations}

With this ordering, the kinetic equation \exref{eq:g} becomes, 
to lowest order in~$\beta_i$,  
\beq
\frac{\dd g}{\dd t} + \frac{\rho_i\vth}{2}\lt\{\la\ephi\ra_\vR,g\rt\} 
= C[g+\la\ephi\ra_\vR F_0] + \frac{2\vpar\la\aext\ra_\vR}{\vth^2}F_0. 
\label{eq:g_low_beta}
\eeq
If we ignore collisions and assume no external forcing ($\aext=0$), 
then $g=0$ is a good solution of this equation (these assumptions 
will be relaxed in \secref{sec:low_beta_partition}). 
The field equations \exsdash{eq:dn}{eq:dB} turn into simple constitutive relations 
\beq
\frac{\dn}{n} = -(1-\hG_0)\ephi,\quad
\frac{\upare}{\vth} = \frac{1}{\beta_i}\hdperp^2\A + \Jext,\quad
\frac{\dB}{B} = \frac{\beta_i}{2}\lt[\frac{Z}{\tau}(1-\hG_0) + (1-\hG_1)\rt]\ephi.
\label{eq:nuB_low_beta}
\eeq
Using the first two of these in \exsdash{eq:A}{eq:cont}, we find that the latter 
become, to lowest order, 
\begin{align}
\label{eq:ZS_A}
&\frac{\dd\A}{\dd t} + \frac{\vth}{2}\dpar\lt[1 + \frac{Z}{\tau}(1-\hG_0)\rt]\ephi 
= \eta\dperp^2\A,\\
&\frac{\rmd}{\rmd t}(1-\hG_0)\ephi = \frac{\vth}{\beta_i}\dpar\hdperp^2\A
+ \vth\dpar\Jext.
\label{eq:ZS_phi}
\end{align}
These equations are the same as those derived by \citet{zocco11} in the limit of 
ultra-low beta ($\beta_e\sim m_e/m_i$), except the electron 
inertia and the coupling to non-isothermal electron kinetics have now been lost---the price 
(painless to pay, in the context of present study, because energetics are not affected) 
for considering somewhat higher~$\beta_e$. 

The system of equations \exsdash{eq:ZS_A}{eq:ZS_phi} 
turns into RMHD \exref{eq:RMHD} when $\kperp\rho_i\ll1$: 
this is shown by using $1-\hG_0\approx -\hdperp^2/2$ [see \exref{eq:G0}]. 
In the opposite limit $\kperp\rho_i\gg1$, using $1-\hG_0\approx 1$, 
one obtains the $\beta\ll1$ limit of the ``Electron RMHD'' equations 
(ERMHD; see S09--\S7.2 or \citealt{boldyrev13}). In the more conventional 
notation involving stream and flux functions [defined in \exref{eq:phi_A_def}], 
they are 
\beq
\frac{\dd\Psi}{\dd t} = \vA\lt(1+\frac{Z}{\tau}\rt)\dpar\Phi + \eta\dperp^2\Psi,\quad
\frac{\dd\Phi}{\dd t} = -\frac{\vA}{2}\dpar\rho_i^2\dperp^2\Psi
\label{eq:ERMHD} 
\eeq  
(we have dropped $\Jext$ because it occurs at large scales).
The relationship between the magnetic field and 
$\Psi$ is still the same as in \exref{eq:PhiPsi_def}. While $\Phi$ is still 
the stream function for the $\vE\times\vB$ velocity, this is now the 
velocity of the electron flow (ions are much slower because of gyroaveraging). 
These equations describe what is sometimes referred to as the turbulence of 
Kinetic Alfv\'en Waves (KAW)---although, like the Alfv\'enic (RMHD) turbulence in the 
inertial range, it is expected to be strong and critically balanced and so does 
not literally consist of waves 
(see S09--\S7.5 and \citealt{cho04,cho09,boldyrev12,tenbarge12,tenbarge13b,boldyrev19}). 

If we consider cold ions, $Z/\tau\gg1$ (but not so cold 
as to break $\beta_e\ll1$), there is an intermediate regime with 
\beq
\frac{Z}{\tau}(1-\Gamma_0) \approx \kperp^2\rhos^2 \sim 1, \quad 
\rhos = \sqrt{\frac{Z}{2\tau}}\,\rho_i = \frac{\cs}{\Omega}, 
\quad \cs = \sqrt{\frac{ZT_e}{m_i}},
\label{eq:rhos_exp}
\eeq
where $\cs$ is the sound speed and $\rhos\gg\rho_i$ is the ``sound radius'', 
setting a transition scale. 
In this regime, the electron-pressure-gradient term [the right-hand side of \exref{eq:cont}] 
is non-negligible and so the AW dynamics become dispersive: using 
\exref{eq:phi_A_def} and \exref{eq:rhos_exp} in \exsdash{eq:ZS_A}{eq:ZS_phi}, 
we arrive at a simple modification of RMHD equations \exref{eq:RMHD} 
\citep[cf.][]{bian09}:
\beq
\frac{\dd\Psi}{\dd t} = \vA\dpar\lt(1 -\rhos^2\dperp^2\rt)\Phi,\quad
\frac{\rmd}{\rmd t}\dperp^2\Phi = \vA\dpar\dperp^2\Psi.
\label{eq:RMHD_rhos}
\eeq 
There is then a second transition in \exsdash{eq:ZS_A}{eq:ZS_phi} 
at $\kperp\rho_i\sim 1$, to ERMHD~\exref{eq:ERMHD}.

\subsection{Linear Theory}
\label{sec:low_beta_lin}

These transitions become particularly transparent if we consider the 
linear dispersion relation for the system \exsdash{eq:ZS_A}{eq:ZS_phi}: 
\beq
\label{eq:omega_low_beta}
\omega^2 = \kpar^2\vA^2\frac{\kperp^2\rho_i^2}{2}
\lt(\frac{1}{1-\Gamma_0} + \frac{Z}{\tau}\rt) 
\approx \lt\{
\begin{array}{ll}
\kpar^2\vA^2(1+\kperp^2\rhos^2),& \kperp\rho_i\ll1,\\\\
\displaystyle
\frac{1+Z/\tau}{2}\,\kpar^2\vA^2\kperp^2\rho_i^2,& \kperp\rho_i\gg1.
\end{array}
\rt.
\eeq
The $\kperp\rho_i\ll1$ limit is the Alfv\'en wave with the dispersive correction due 
to the electron-pressure gradient.  
The $\kperp\rho_i\gg1$ limit is the KAW dispersion relation with $\beta\ll1$ (see S09--\S7.3). 
When $Z/\tau\gg1$, it becomes $\omega^2\approx\kpar^2\vA^2\kperp^2\rhos^2$ and so 
the transition between the long- and short-wavelength frequencies 
is seamless. Thus, in this limit, the transition between the AW and KAW cascades 
occurs at $\kperp\rhos\sim1$.    

\subsection{Free-Energy Budget}
\label{sec:W_low_beta}

The nonlinear system \exsdash{eq:ZS_A}{eq:ZS_phi} has a conserved energy 
\beq
W = \frac{\vth^2}{4}\int\frac{\rmd^3\vr}{V}\lt[\ephi(1-\hG_0)\ephi 
+ \frac{Z}{\tau}\bigl|(1-\hG_0)\ephi\bigr|^2 + \frac{2}{\beta_i}\bigl|\hdperp\A\bigr|^2\rt], 
\label{eq:W_low_beta}
\eeq
which is the appropriate low-beta, $g=0$ limit of \exref{eq:W_g}. 
Note that, whereas $\dn/n$ does appear in \exref{eq:W_low_beta}
(the second term), $\dB$ is energetically (and dynamically) insignificant 
[see \exref{eq:nuB_low_beta}]. 

At $\kperp\rhos\ll1$, \exref{eq:W_low_beta} reduces to the energy of Alfv\'en waves 
\beq
\WA = \frac{1}{2}\int\frac{\rmd^3\vr}{V}\lt(|\vdperp\Phi|^2 + |\vdperp\Psi|^2\rt) 
= \frac{1}{2}\int\frac{\rmd^3\vr}{V}\lt(|\vuperp|^2 + |\vbperp|^2\rt), 
\label{eq:WA}
\eeq
conserved by RMHD \exref{eq:RMHD}. 
When $\kperp\rhos\sim 1$ but $\kperp\rho_i\ll1$, 
\beq
W = \frac{1}{2}\int\frac{\rmd^3\vr}{V}\lt(|\vdperp\Phi|^2 + 
\rhos^2|\dperp^2\Phi|^2 + |\vdperp\Psi|^2\rt), 
\label{eq:W_rhos}
\eeq
the energy of the system \exref{eq:RMHD_rhos}. At $\kperp\rho_i\gg1$, $W$ becomes 
the energy of low-beta KAW perturbations described by \exref{eq:ERMHD}:\footnote{Note 
the typo in S09--\S7.8, where this is derived: a missing factor of 2 in front 
of $\Phi^2$ in Eq.~(246).} 
\beq
\WK = \int\frac{\rmd^3\vr}{V}
\lt[\lt(1+\frac{Z}{\tau}\rt)\frac{\Phi^2}{\rho_i^2} + \frac{1}{2}|\vdperp\Psi|^2\rt].
\label{eq:WK}
\eeq

The existence of the invariant \exref{eq:W_low_beta}, valid uniformly 
at small, order-unity and large $\kperp\rho_i$,  
means that no damping of anything and, therefore, no ion heating 
occurs at any wave number, until resistivity kicks in and causes electron heating: 
it is easy to ascertain that 
\beq
\frac{\rmd W}{\rmd t} = \epsA - Q_e,
\eeq 
where $\epsA$ is given by \exref{eq:epsA_def} and $Q_e$ by \exref{eq:Qe_def}. 
In steady state, $Q_e=\epsA$. 

\subsection{Energy Partition in the Presence of Compressive Cascade}
\label{sec:low_beta_partition}

In the above, we assumed the $g=0$ solution for the kinetic equation \exref{eq:g_low_beta}. 
This corresponds to a situation in which only Alfv\'enic perturbations are stirred 
up at the largest scales: indeed, the relations \exref{eq:nuB_low_beta} imply that the 
compressive fields $\dn$ and $\dB$ peter out at $\kperp\rho_i\ll1$. 
Let us now relax this assumption. Mathematically, this would correspond, e.g., 
to restoring the external parallel acceleration term 
in \exref{eq:g_low_beta}. 
The variance of the forced kinetic scalar described by \exref{eq:g_low_beta} with $\aext\neq0$
is conserved by the nonlinearity: 
\beq
\frac{\rmd}{\rmd t}\frac{\vth^2}{4}\int\frac{\rmd^3\vr}{V}\overline{\frac{\la g^2\ra_\vr}{F_0}} 
- \frac{\vth^2}{2}
\int\frac{\rmd^3\vr}{V}\overline{\lt\la\frac{g C[g + \la\ephi\ra_\vR F_0]}{F_0}\rt\ra_\vr}
= \int\frac{\rmd^3\vr}{V}\overline{\aext\la\vpar g\ra_\vr} = \epsC, 
\label{eq:gsq_ev}
\eeq
where $\epsC$ is the energy flux in the compressive cascade [cf.~\exref{eq:epsC_def}]. 
In steady state ($\rmd/\rmd t=0$), 
we have a balance between this compressive input power and the collisional terms [cf.~\exref{eq:W_evo}]: 
\beq
\label{eq:compr_bal}
\epsC = Q_i + \Qx,\quad
\Qx = \frac{\vth^2}{2}\int\frac{\rmd^3\vr}{V}\overline{\lt\la\la\ephi\ra_\vR C[h]\rt\ra_\vr}
= \frac{\vth^2}{2}\int\frac{\rmd^3\vr}{V}\ephi \overline{\la C[h]\ra_\vr},
\eeq 
where $Q_i$ is given by \exref{eq:Qi_def}.
Thus, all the compressive energy becomes ion heat, with the exception 
of the collisional energy exchange $\Qx$ with Alfv\'enic perturbations, 
which is, as we are about to argue, small when collisions are weak. 
The implication is that all the Alfv\'enic energy is destined, via 
the AW cascade smoothly transitioning into the KAW cascade, to be dissipated 
into electron heat, 
\beq
\epsA = \epsK = Q_e.
\eeq
We will confirm this directly in \secref{sec:AW_coupled}. 

If the collision frequency is small compared to the forcing or 
nonlinear-advection time scales in \exref{eq:g_low_beta}, the only way for 
the collision terms to balance the finite energy flux is for $g$ to develop 
small scales in phase space, thus activating large derivatives in $C[h]$. 
This is indeed what happens, as the nonlinear term successfully pushes 
$g$ towards small scales in both $\vR$ and $\vperp$, viz., 
towards $\dvperp/\vth \sim (\kperp\rho_i)^{-1}\ll1$, 
via a process known as the entropy cascade (see S09--\S7.9). 
This is a route to ion heating that requires no parallel streaming 
and is, therefore, the only feasible one 
for the effectively 2D kinetic equation \exref{eq:g_low_beta}, 
where the parallel streaming has been ordered out due to low $\beta_i$ 
\citep[cf.][]{tatsuno09,plunk10}. Recent numerical results 
by \citet{kawazura19} appear to confirm the presence of such an ion-heating 
route in low-beta GK turbulence. 

The ion heating rate $Q_i$ [see \exref{eq:Qi_def}] is positive definite 
and by this process it will be rendered finite, i.e., independent of the ion collision rate, 
however small the latter is. 
Let us estimate the size of $\Qx$ in comparison to $Q_i$. 
Clearly, only the parts of $\ephi$ and $h$ that vary on fine scales in position 
and velocity space matter in $\Qx$ and $Q_i$, the contribution from large scales  
being small because the collision frequency is small. 
The GK collision operator is a diffusion operator both in velocity and position  
\citep[see, e.g.,][]{abel08}, with the size of the position and velocity gradients 
comparable in the entropy cascade. At $\kperp\rho_i\gg1$, when collisions become important, 
\begin{align}
\label{eq:Qi_est}
Q_i &\sim \vth^2 \nu_{ii} (\kperp\rho_i)^2 \frac{h^2}{F_0^2},\\
\Qx &\sim \vth^2 \nu_{ii} (\kperp\rho_i)^{3/2} \frac{h}{F_0}\ephi 
\sim \vth^2 \nu_{ii} \kperp\rho_i \frac{h^2}{F_0^2},
\label{eq:Qx_est}
\end{align}
where $\nu_{ii}$ is the ion collision frequency. Thus, $\Qx\ll Q_i$. 
Here $\Qx$ loses out compared to $Q_i$ by one factor of $(\kperp\rho_i)^{1/2}$ 
because of the gyroaveraging under the velocity integral of $C[h]$ and 
by another factor of $(\kperp\rho_i)^{1/2}$ because, as will be evident 
from \exsdash{eq:A_coupled}{eq:dn_coupled}, we must order 
$\ephi \sim \overline{\hJ_0 g} \sim (\kperp\rho_i)^{-1/2}h/F_0$ 
in order for the compressive perturbations 
to have any relevance. In fact, \exref{eq:Qx_est} is probably an overestimate 
because $\Qx$ is not sign-definite and so there will also be a tendency 
for the small-scale variation within it to average out under integration. 
In any event, it is clear that when collisions are weak, the collisional 
energy exchange can be neglected. 

\subsection{Effect of Compressive Cascade on Alfv\'enic Cascade}
\label{sec:AW_coupled}

For completeness, let us ascertain that the notion that non-zero $g$ has no 
energetic effect on the AW and KAW cascades is consistent with the dynamical 
equations for the latter. We allow $g/F_0 \sim \ephi$ as per \exref{eq:low_beta_order}. 
In this case, $\overline{\vpar\hJ_0 g}$ is still one-order subdominant in 
\exref{eq:upar} and $\dB/B$ is still small compared to $\dn/n$, but 
there is now a contribution from $g$ to $\dn/n$ in \exref{eq:dn}. 
The resulting pair of equations, replacing \exsdash{eq:ZS_A}{eq:ZS_phi},~is 
\begin{align}
\label{eq:A_coupled}
&\frac{\dd\A}{\dd t} 
+ \frac{\vth}{2}\dpar\lt\{\ephi 
+ \frac{Z}{\tau}\lt[(1-\hG_0)\ephi - \overline{\hJ_0 g}\rt]\rt\} = \eta\dperp^2\A,\\
&\frac{\rmd}{\rmd t}\lt[(1-\hG_0)\ephi - \overline{\hJ_0 g}\rt] 
- \frac{\vth}{\beta_i}\dpar\hdperp^2\A = \vth\dpar\Jext, 
\label{eq:dn_coupled}
\end{align}
coupled to \exref{eq:g_low_beta}. 

The quantity in the square brackets 
in \exref{eq:A_coupled} and \exref{eq:dn_coupled} is $-\dn/n$, so 
these equations can be thought of as evolution equations of $\A$ and $\dn/n$, 
the latter's relationship to $\ephi$ now involving~$g$. 
Alternatively, \exref{eq:dn_coupled} can be recast as 
\beq
\frac{\rmd}{\rmd t}(1-\hG_0)\ephi - \frac{\vth}{\beta_i}\dpar\hdperp^2\A 
= \vth\dpar\Jext
-\frac{\rho_i\vth}{2}\overline{\lt\la\lt\{\la\ephi\ra_\vR-\ephi,g\rt\}\rt\ra_\vr} 
+ \overline{\lt\la C[g+\la\ephi\ra_\vR F_0]\rt\ra_\vr} 
\label{eq:phi_coupled}
\eeq
if one uses the evolution equation for $\overline{\hJ_0 g}$ derived 
by integrating \exref{eq:g_low_beta} over the velocity space [\exref{eq:phi_coupled} 
can also be obtained by applying the ordering \exref{eq:low_beta_order} 
to \exref{eq:gint}].\footnote{Our choice of forcing in \exref{eq:g_low_beta} 
has ensured that the contribution of $g$ to density is not affected and so 
the compressive driving does not stir up Alfv\'enic perturbations.} 
This emphasises the nonlinear FLR coupling of $\ephi$ to $g$. 

These equations support a generalised version of the (collisionless) 
invariant~\exref{eq:W_low_beta}:
\beq
\tW = \frac{\vth^2}{4}\int\frac{\rmd^3\vr}{V}\lt[\ephi(1-\hG_0)\ephi 
+ \frac{Z}{\tau}\lt|(1-\hG_0)\ephi - \overline{\hJ_0 g}\rt|^2 
+ \frac{2}{\beta_i}\bigl|\hdperp\A\bigr|^2\rt],
\label{eq:W_coupled}
\eeq
which is the low-beta limit of \exref{eq:W_g}, excluding the variance of $g$, which is 
still conserved independently [see \exref{eq:gsq_ev}].
Indeed, using \exref{eq:phi_coupled} to work out the time derivative of the first term and 
\exref{eq:A_coupled} and \exref{eq:dn_coupled} for the other two terms, we get 
\beq
\frac{\rmd\tW}{\rmd t} = \epsA - Q_e + \Qx,
\label{eq:W_ev}
\eeq
where $\epsA$ is given by \exref{eq:epsA_def}, $Q_e$ by \exref{eq:Qe_def} 
and $\Qx$ in \exref{eq:compr_bal}. 
The nonlinear terms have vanished by cancellation and because 
\beq
\int\frac{\rmd^3\vr}{V}\ephi\overline{\lt\la\lt\{\la\ephi\ra_\vR-\ephi,g\rt\}\rt\ra_\vr} 
= \overline{\int\frac{\rmd^3\vR}{V}\la\ephi\ra_\vR \lt\{\la\ephi\ra_\vR,g\rt\}} = 0 
\eeq
(after swapping the order of the $\vv$ and $\vr$ integration and changing 
the integration variable from $\vr$ to $\vR$). 

Combining \exref{eq:W_ev} 
and \exref{eq:gsq_ev}, we recover the overall conservation law \exref{eq:W_evo}, 
as indeed we must, because the free energy is 
\beq
W = \tW + \frac{\vth^2}{4}\int\frac{\rmd^3\vr}{V}\overline{\frac{\la g^2\ra_\vr}{F_0}}. 
\eeq
However, we now have more restrictive and, therefore, more informative energy balances 
\exref{eq:compr_bal} and \exref{eq:W_ev} (with $\rmd\tW/\rmd t = 0$ in steady state). 
Since, as we argued in \secref{sec:low_beta_partition}, $\Qx$ is small, we conclude that 
\beq
Q_i = \epsC, \quad Q_e = \epsA,
\label{eq:QiQe}
\eeq 
so compressive energy goes into ions, Alfv\'enic into electrons.  
Thus, while non-zero $g$ does insinuate itself into the dynamics of Alfv\'enic perturbations, 
there is no energy exchange between the two cascades. 

\subsection{Ultra-Low Beta}
\label{sec:ultra_low}

Formally, there is an interesting very-low-beta limit that is outside the validity 
of our theory so far. Namely, if $\beta_e\sim m_e/m_i$, we can no longer use the 
isothermal-electron-fluid approximation introduced in \secref{sec:IEF}. 
The equations in this case are quite similar to \exref{eq:g_low_beta} 
and \exsdash{eq:A_coupled}{eq:dn_coupled}, except in \exref{eq:A_coupled} 
there is now an electron-inertia term and a piece of parallel pressure gradient 
that contains a non-zero parallel electron temperature perturbation. 
The latter has to be calculated from the electron drift-kinetic equation, 
thus opening up an electron heating route via parallel heat transport and 
Landau damping. With $g=0$, the appropriate equations were worked out by 
\citet{zocco11} and proved to be a useful model for numerical experimentation 
\citep{loureiro13colless,loureiro16viriato,groselj17}; 
they can be generalised to $g\neq 0$ 
in exactly the same way as the system \exsdash{eq:ZS_A}{eq:ZS_phi} was generalised 
in \secsand{sec:low_beta_partition}{sec:AW_coupled}. 
There is no change in the energy partition: by the same arguments as above, 
the energy of compressive perturbations goes into ions, the energy of Alfv\'enic 
ones into electrons. 

\section{Reduced Dynamics and Heating in the Hall Limit} 
\label{sec:Hall}

Let us now consider the case of $\beta_e\sim1$ and $\beta_i\ll1$. 
This is the so-called Hall limit and the derivation 
in \secsdash{sec:Hall_ordering}{sec:Hall_lin} 
is a reworking (in a slightly different order) 
of the ``Hall RMHD'' (S09--\S{E}), which we will need for what follows 
and which turns out to have some interesting consequences for the energy partition, 
detailed in \secsand{sec:KAW_ICW}{sec:Hall_Larmor}.  

\subsection{Ordering}
\label{sec:Hall_ordering}
 
In this limit, since $\beta_e = Z\beta_i/\tau$, the ions are cold and, as we anticipate  
based on \secref{sec:low_betae}, the AW physics will become dispersive at $\kperp\rhos\sim1$:  
\beq
\frac{Z}{\tau} \sim \frac{1}{\beta_i}\gg1,\quad
\kperp\rhos\sim1 \hence
\kperp\rho_i \sim \sqrt{\frac{\tau}{Z}} \sim \sqrt{\beta_i} \ll1
\hence  
\kperp d_i \sim 1,
\label{eq:Hall_order}
\eeq
where $d_i=\rho_i/\sqrt{\beta_i} = \rhos\sqrt{2/\beta_e}$ 
is the ion inertial scale, which is of the same 
order as $\rhos$ in this limit. 

We must adjust all expansions and equations accordingly. 
Instead of \exref{eq:A_order} and \exref{eq:phi_order}, 
we have 
\beq
\A \sim \frac{\epsilon}{\kperp\rho_i} \sim \frac{\epsilon}{\sqrt{\beta_i}},
\quad
\ephi \sim \frac{\epsilon}{\kperp\rho_i\sqrt{\beta_i}} \sim \frac{\epsilon}{\beta_i}.
\label{eq:Hall_A_phi_order}
\eeq
Since the sound speed and the Alfv\'en speed are of the same order in this limit, viz., 
\beq
\cs = \sqrt{\frac{ZT_e}{m_i}} = \vth \sqrt{\frac{Z}{2\tau}}
= \vA\sqrt{\frac{\beta_e}{2}} \sim \vA,
\eeq 
the AW and the compressive modes (slow waves) have similar frequencies. 
This allows us to handle both cascades simultaneously. To avoid prejudice, 
we order the compressive perturbations to have similar amplitudes to the 
Alfv\'enic ones: 
\beq
\frac{\dn}{n} \sim \frac{\upari}{\vA} \sim \frac{\dB}{B} \sim 
\frac{\dBperp}{B}\sim\epsilon, 
\label{eq:Hall_n_order}
\eeq 
where $\upari = \overline{\vpar\hJ_0 g}$ is the parallel ion flow velocity. 
The requirement that \exsdash{eq:dn}{eq:dB} be consistent with \exref{eq:Hall_n_order}
implies that we ought to order 
\beq
\frac{g}{F_0} \sim \frac{\epsilon}{\sqrt{\beta_i}},\quad 
\overline{\hJ_0 g} \sim \epsilon,
\label{eq:Hall_g_order}
\eeq
i.e., to lowest order, the distribution function should have no density moment.  

\subsection{Equations}
\label{sec:Hall_eqns}

With these orderings, \exsdash{eq:dn}{eq:dB} become, to lowest order in $\beta_i$ (and $\tau$),  
\beq
\overline{g} = \frac{\dn}{n} - \frac{1}{2}\hdperp^2\ephi,\quad 
\upare = \upari + \vth\lt(\frac{1}{\beta_i}\hdperp^2\A + \Jext\rt),\quad
\frac{\dn}{n} = -\frac{2}{\beta_e}\frac{\dB}{B}.
\label{eq:Hall_fields}
\eeq
The last of these equations is the balance between the magnetic and electron pressure, 
ions being too cold to matter. Using this relationship in \exsdash{eq:A}{eq:cont}, 
we get\footnote{The resistive term in \exref{eq:Hall_A} can, in fact, 
be legitimately retained only if resistivity becomes important 
before the Larmor scale is reached. This is possible formally, but unlikely in reality.} 
\begin{align}
\label{eq:Hall_A}
&\frac{\dd\A}{\dd t} 
+ \frac{\vth}{2}\dpar\lt(\ephi + \frac{2}{\beta_i}\frac{\dB}{B}\rt) = 
\eta\dperp^2\A,\\
&\lt(1+\frac{2}{\beta_e}\rt)\frac{\rmd}{\rmd t}\frac{\dB}{B} = 
\dpar\lt[\upari + \vth\lt(\frac{1}{\beta_i}\hdperp^2\A + \Jext\rt)\rt].
\label{eq:Hall_dB}
\end{align}
So all four fields $\A$, $\ephi$, $\dB$ and $\upari$ (the latter representing $g$) 
are coupled and we need two more equations to close the system. 

One of these is \exref{eq:gint}, where applying the ordering of \secref{sec:Hall_ordering} 
leads to the disappearance of the entire right-hand side, as well as of the $\dB$ term 
under the time derivative. To lowest order, therefore, we are left with a rather 
familiar equation [cf.~the second RMHD equation in \exref{eq:RMHD}]:
\beq
\frac{\rmd}{\rmd t} \frac{1}{2}\hdperp^2\ephi + \frac{\vth}{\beta_i}\dpar\hdperp^2\A = 
-\vth\dpar\Jext. 
\label{eq:Hall_phi}
\eeq

The last required equation is the lowest-order version of the kinetic equation~\exref{eq:g}:  
\beq
\frac{\rmd g}{\rmd t} = \vpar\dpar\frac{2}{\beta_i}\frac{\dB}{B} F_0 + C[g] 
+ \frac{2\vpar\aext}{\vth^2}F_0, 
\label{eq:Hall_g}
\eeq 
where we again used the last equation in \exref{eq:Hall_fields}. 
This is consistent with $\overline{g} = 0$ to lowest order, as anticipated in 
\exref{eq:Hall_g_order}. If we split off the velocity moment from $g$, viz., 
\beq
g = \frac{2\upari\vpar}{\vth^2} F_0 + \G,\quad \overline{\G} = 0,\quad 
\overline{\vpar\G} = 0, 
\eeq 
then \exref{eq:Hall_g} becomes 
\begin{align}
\label{eq:Hall_upar}
\frac{\rmd\upari}{\rmd t} &= \vA^2\dpar\frac{\dB}{B} + \aext,\\
\frac{\rmd\G}{\rmd t} &= C[\G]. 
\label{eq:Hall_G}
\end{align}
The first of these is the final equation that we needed to close the system 
comprising already \exref{eq:Hall_A}, \exref{eq:Hall_dB} and \exref{eq:Hall_phi}. 
The second equation, \exref{eq:Hall_G}, 
describes a passively advected kinetic field, which, however, 
is not coupled to anything and so can be safely put to zero\footnote{Unless 
it is explicitly forced. The forcing that we have chosen for compressive 
perturbations has ended up only driving parallel ion flows. To model energy injection 
into $\G$, we would need to inject, e.g., temperature perturbations---physically 
this can happen if there is an equilibrium temperature gradient 
(see, e.g., \citealt{sch16} or \citealt{xukunz16}), 
but we shall not consider such equilibria here. \label{fn:strat}}---it is 
the kinetic version of the MHD entropy mode, whereas the rest of our equations describe 
linearly and nonlinearly coupled AW and slow waves (SW). Note finally that 
\exref{eq:Hall_phi} is needed because it is not possible to calculate $\ephi$ from 
the first and last of the field equations \exref{eq:Hall_fields} and the 
kinetic equation \exref{eq:Hall_g}. This is because, as assumed in \exref{eq:Hall_g_order}, 
the density moment $\overline{g}$ comes 
from the next-order part of $g$ not captured in \exref{eq:Hall_g}. 

It is instructive to rewrite the Hall RMHD equations \exsdash{eq:Hall_A}{eq:Hall_phi}
and \exref{eq:Hall_upar} in ``fluid'' notation, dropping the forcing terms and resistivity 
\citep[cf.][]{gomez08}: 
\begin{align}
\label{eq:Hall_Psi}
&\frac{\dd\Psi}{\dd t} = \vA\dpar\bigl(\Phi + \vA \rhoh\B\bigr),\\
\label{eq:Hall_B}
&\frac{\rmd\B}{\rmd t} = \dpar\bigl(\vs\,\U - \rhoh\dperp^2\Psi\bigr),\\
\label{eq:Hall_Phi}
&\frac{\rmd}{\rmd t}\dperp^2\Phi = \vA\dpar\dperp^2\Psi,\\
\label{eq:Hall_U}
&\frac{\rmd\,\U}{\rmd t} = \vs\dpar\B,
\end{align} 
where $\Phi$ and $\Psi$ are defined by \exref{eq:phi_A_def}, 
we have denoted 
\beq
\label{eq:BU_def}
\B = \frac{\dB}{B}\sqrt{1+\frac{2}{\beta_e}},\quad
\U = \frac{\upari}{\vA},
\eeq
and introduced the Hall transition scale 
\beq
\label{eq:rhoh_def}
\rhoh = \frac{d_i}{\sqrt{1+2/\beta_e}} = \frac{\rhos}{\sqrt{1+\beta_e/2}} 
= \rho_i \sqrt{\frac{Z/\tau}{2+\beta_e}}
\eeq
and the SW phase speed 
\beq
\vs = \frac{\vA}{\sqrt{1+2/\beta_e}} = \frac{\cs}{\sqrt{1+\beta_e/2}}.
\eeq

At $\kperp\rhoh\ll1$, the Alfv\'enic and the SW-like perturbations 
decouple from each other and revert to standard RMHD equations (see S09--\S2.4): 
the AW equations \exref{eq:Hall_Psi} and \exref{eq:Hall_Phi} become \exref{eq:RMHD}  
and the SW equations \exref{eq:Hall_B} and \exref{eq:Hall_U} become 
\beq
\frac{\rmd\B}{\rmd t} = \vs\dpar \U,\quad
\frac{\rmd\,\U}{\rmd t} = \vs\dpar\B
\label{eq:Hall_SW}
\eeq
(passively advected by the AW via $\rmd/\rmd t$ and $\dpar$, without energy exchange). 
Thus, our new system of equations \exsdash{eq:Hall_Psi}{eq:Hall_U} captures the RMHD 
regime and describes its transformation, at the Hall transition scale $\rhoh$, into 
one in which all four fields $\Phi$, $\Psi$, $\B$ and $\U$ are coupled. 

The system \exsdash{eq:Hall_Psi}{eq:Hall_U} also contains the low-$\beta_e$  
limit~\exref{eq:RMHD_rhos}. This corresponds to taking the limit $\vs\to 0$. 
Combining \exref{eq:Hall_B} and \exref{eq:Hall_Phi}, we get 
\beq
\frac{\rmd}{\rmd t}\lt(\B + \frac{\rhoh}{\vA}\dperp^2 \Phi\rt) = \vs\dpar\U \to 0
\hence \B = -\frac{\rhoh}{\vA}\dperp^2 \Phi.
\label{eq:Hall_low_betae}
\eeq
Using this in \exref{eq:Hall_Psi} and setting $\rhoh=\rhos$, we get the first 
equation in~\exref{eq:RMHD_rhos}. The second is the same as~\exref{eq:Hall_Phi}. 
The parallel velocity in this limit decouples and cascades independently: 
\beq
\label{eq:U_passive}
\frac{\rmd\,\U}{\rmd t} = 0,
\eeq
just like $\G$ does in \exref{eq:Hall_G} and like $g$ did in~\exref{eq:g_low_beta}. 

\subsection{Free Energy and Heating}
\label{sec:W_Hall}

The conserved free energy for \exsdash{eq:Hall_Psi}{eq:Hall_U} 
[equivalently, for \exsdash{eq:Hall_A}{eq:Hall_phi} and \exref{eq:Hall_upar}]~is
\begin{align}
\nonumber
\tW &= \frac{1}{2}\int\frac{\rmd^3\vr}{V}\lt[\bigl|\dperp\Phi\bigr|^2 + \bigl|\dperp\Psi\bigr|^2 
+ \vA^2\lt(\U^2 + \B^2\rt)\rt]\\
&= \int\frac{\rmd^3\vr}{V}\lt[\frac{\vth^2}{4}\lt(
\frac{1}{2}\bigl|\hdperp\ephi\bigr|^2 
+ \frac{2}{\beta_i}\bigl|\hdperp\A\bigr|^2\rt) 
+ \frac{\upari^2}{2}
+ \frac{\dB^2}{8\pi m_i n_i}\lt(1+\frac{2}{\beta_e}\rt)\rt].
\label{eq:W_Hall}
\end{align}
The free energy has no access to~$\G$, whose variance is individually conserved, 
as is obvious from \exref{eq:Hall_G}. If we forced $\G$ (without breaking 
the ordering of \secref{sec:Hall_ordering}), the free energy injected in this way would remain 
decoupled and travel all through the Hall range of scales unconcerned with the wave dynamics, 
eventually arriving at $\kperp\rho_i\sim1$ and transiting into the sub-Larmor-scale 
ion entropy cascade and eventually into ion heat (see \secref{sec:Hall_Larmor_ions}).
As we already mentioned in footnote~\ref{fn:strat}, a natural physical way in which $\G$ 
might be forced is by the presence of an ion temperature gradient. However, there cannot 
be net heating of the plasma by turbulence produced by temperature gradients: any 
energy thus ``borrowed'' from the ion thermal bath may only be redistributed between 
species \citep{abel13}. In the present case, all of it is destined for ions.  

The first two terms in \exref{eq:W_Hall} are the Alfv\'enic energy \exref{eq:WA}, 
conserved by the RMHD \exref{eq:RMHD}; the last two terms are the slow-wave energy 
\beq
\WS = \frac{\vA^2}{2}\int\frac{\rmd^3\vr}{V}\lt(\U^2 + \B^2\rt),
\eeq
conserved by \exref{eq:Hall_SW}. 
When $\beta_e\ll1$, the substitution of \exref{eq:Hall_low_betae} turns 
\exref{eq:W_Hall} into \exref{eq:W_rhos}, with the $\U^2$ part of the free energy 
splitting off, destined for ion heating. In contrast, at $\beta_e\sim 1$, 
the decoupling between the Alfv\'enic and compressive cascades is broken 
at $\kperp\rhoh\sim1$, so we can no longer conclude that the former must 
heat electrons and the latter ions.  
In order to work out what happens (see \secref{sec:Hall_heating} for a preview of the answer), 
we must shift our focus to $\kperp\rho_i\sim1$ (\secref{sec:Hall_Larmor}), 
but for that, we must first investigate into what 
kind of turbulence the Hall turbulence turns at $\kperp\rhoh\gg1$ (\secref{sec:KAW_ICW}). 
In working this out, we will find linear theory to be a valuable guide. 

\subsection{Linear Theory}
\label{sec:Hall_lin}

The dispersion relation~is 
\beq
\label{eq:Hall_DR}
(\omega^2 - \oA^2)(\omega^2-\oS^2) = \omega^2\oK^2, 
\eeq
where $\oA = \kpar\vA$ is the AW frequency, $\oS=\kpar \vs$ the SW frequency 
and $\oK = \kpar\vA\kperp\rhoh$ the KAW frequency in the $Z/\tau\gg1$ limit (cf.\ S09--\S7.3). 
There is no damping of anything here because ions cannot stream along the 
field lines as fast as waves propagate [see \exref{eq:Hall_g}]. 

At $\kperp\rhoh\ll1$, $\oK\ll\oA,\oS$ and 
we recover from \exref{eq:Hall_DR} four low-frequency MHD waves
\beq
\omega=\pm\oA,\quad \omega=\pm\oS. 
\eeq
At $\kperp\rhoh\gg1$, if $\omega\gg\oA,\oS$,  
the linear response assumes its KAW form:\footnote{As it did in the 
$\beta_e\ll1$ limit treated in \secref{sec:low_beta_lin}. It is also not hard to see that, 
at $\beta_e\ll1$, $\oS\ll\oA$ and the Alfv\'enic branch in \exref{eq:Hall_DR} 
obeys the $\kperp\rho_i\ll1$ version of \exref{eq:omega_low_beta}.}
\beq
\omega = \pm\oK = \pm \kpar\vA\kperp\rhoh. 
\eeq 
This is not particularly surprising: the KAW response is the Alfv\'enic response 
with (nearly) immobile ions---and the ion-flow terms in the two magnetic-field equations 
\exref{eq:Hall_Psi} and \exref{eq:Hall_B} do indeed become subdominant at $\kperp\rhoh\gg1$. 
Linearly, the KAW are then described by 
\beq
\frac{\dd\Psi}{\dd t} = \vA^2\rhoh\frac{\dd\B}{\dd z},\quad
\frac{\dd\B}{\dd t} = -\rhoh\frac{\dd}{\dd z}\dperp^2\Psi.
\label{eq:KAW_lin}
\eeq
In the Hall limit, there is nothing particularly kinetic about kinetic Alfv\'en waves, 
so they should probably be called Hall Alfv\'en waves (but are sometimes called whistlers); 
we shall keep the KAW moniker to avoid multiplying entities beyond necessity.

There is more to the story at $\kperp\rhoh\gg1$. In this limit, 
besides the two KAW, \exref{eq:Hall_DR} has two other, low-frequency, solutions:
\beq
\omega = \pm\frac{\oA\oS}{\oK} = \pm \frac{\kpar\vs}{\kperp\rhoh} 
= \pm\Omega\,\frac{\kpar}{\kperp} \equiv \pm\oI. 
\label{eq:ICW}
\eeq 
These are oblique ion cyclotron waves (ICW; cf.~\citealt{sahraoui07}). 
For these perturbations, \exref{eq:Hall_Psi} and \exref{eq:Hall_B}
become quasistatic, viz.,   
\beq
\B = - \frac{\Phi}{\vA\rhoh},\quad
\dperp^2\Psi = \frac{\vs}{\rhoh}\,\U = \Omega\,\U, 
\label{eq:ICW_lin_const}
\eeq
and, consequently, the linearised versions of 
\exref{eq:Hall_Phi} and \exref{eq:Hall_U} turn into
\beq
\frac{\dd}{\dd t}\dperp^2\Phi = \Omega\,\frac{\dd\upari}{\dd z},\quad
\frac{\dd\upari}{\dd t} = -\Omega\,\frac{\dd\Phi}{\dd z}.
\label{eq:ICW_lin}
\eeq
It is more transparent here to go back from $\U$ [defined in \exref{eq:BU_def}] to $\upari$ 
as the Alfv\'enic normalisation is no longer physically relevant. 
These equations, and the corresponding dispersion relation \exref{eq:ICW}, 
are mathematically the same as the equations and the dispersion relation 
for inertial waves in rigidly rotating (with angular velocity $\Omega/2$) 
neutral fluids \citep[see, e.g.,][]{nazarenko11,davidson13}. 
We shall see momentarily that the analogy survives also nonlinearly 
and that, therefore, ICW turbulence displays some familiar features. 

\begin{figure}
\centerline{\includegraphics[width=0.45\textwidth]{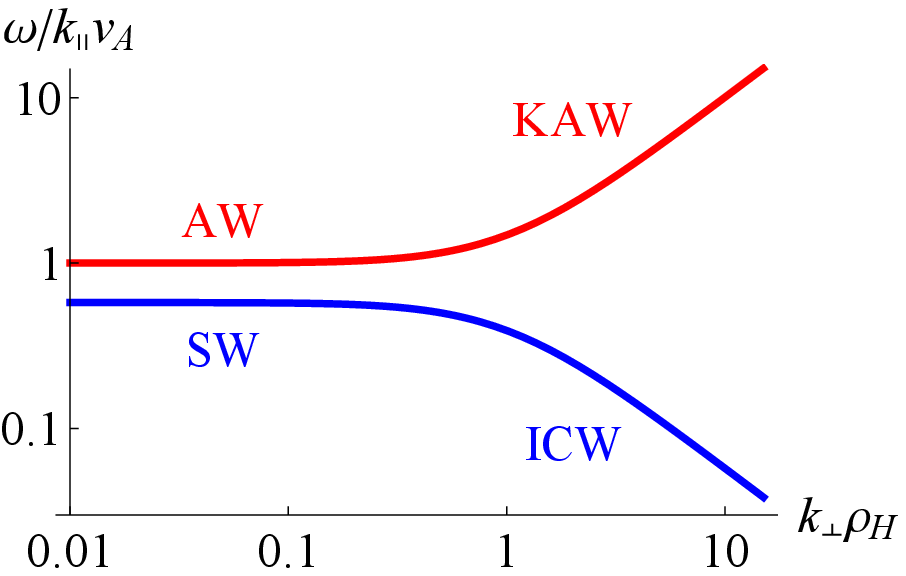}} 
\vskip2mm
\caption{Solutions \exref{eq:Hall_waves} of the Hall dispersion relation \exref{eq:Hall_DR} 
with $\beta_e=1$.}
\label{fig:waves}
\end{figure}

Finally, \figref{fig:waves} shows the full solutions of \exref{eq:Hall_DR}, 
\beq
\label{eq:Hall_waves}
\omega^2 = \frac{\kpar^2\vA^2}{2}\lt\{1 + \sigma^2 + \kperp^2\rhoh^2 \pm 
\sqrt{\bigl[\kperp^2\rhoh^2 + (1+\sigma)^2\bigr]\bigl[\kperp^2\rhoh^2 + (1-\sigma)^2\bigr]}\rt\},
\eeq
where 
\beq
\sigma = \frac{\vs}{\vA} = \frac{1}{\sqrt{1+2/\beta_e}}
\label{eq:sigma_def}
\eeq
(the only parameter in the problem). Note that there is no mode conversion, 
the AW continuously turn into KAW and SW into~ICW. 
The two curves separate ever further at smaller~$\beta_e$, tending 
towards the limit described by~\exref{eq:RMHD_rhos} and \exref{eq:Hall_low_betae}.

\subsection{Hall Turbulence at Short Wavelengths} 
\label{sec:KAW_ICW}

The nature of Hall turbulence at $\kperp\rhoh\gg1$ is determined by the the way in which
fast (KAW) and slow (ICW) perturbations interact with themselves and (potentially) with each other. 

We are dealing with a two-time-scale problem, so let us split all our fields 
into slow and fast components, with ``slow'' defined as the average of the
relevant field over the KAW period and ``fast'' as the difference between 
that and the exact field: 
\beq
\Psi = \bPsi + \tPsi,\quad
\Phi = \bPhi + \tPhi,\quad
\B = \bB + \tB,\quad
\U = \bU + \tU. 
\label{eq:field_split}
\eeq 
In everything that follows, overbar will mean KAW-time-scale averaging and 
overtilde will designate KAW-time-scale quantities, which average to zero 
(with apologies to the reader, who should now forget what 
overbars and overtildes have been used for previously). 
The slow quantities will represent the ICW turbulence and the fast ones 
the KAW turbulence. 

We shall venture an {\em a priori} guess that the two cascades will decouple 
completely at $\kperp\rhoh\gg1$, work out the scalings of all the fields 
on that basis and then confirm {\em a posteriori} that those are consistent 
with such a decoupling. Namely, we anticipate that the nonlinear version 
of the KAW equations \exref{eq:KAW_lin} will be 
\beq
\frac{\dd\tPsi}{\dd t} = \vA^2\rhoh\widetilde{\dpar\tB},\quad
\frac{\dd\tB}{\dd t} = -\rhoh\widetilde{\dpar\dperp^2\tPsi},\quad
\dpar = \frac{\dd}{\dd z} + \frac{1}{\vA}\{\tPsi,\dots\}, 
\label{eq:KAW_nlin}
\eeq
and the nonlinear version of the ICW equations \exref{eq:ICW_lin}
\beq
\frac{\rmd}{\rmd t}\dperp^2\bPhi = \Omega\vA\frac{\dd\,\bU}{\dd z},\quad
\frac{\rmd\,\bU}{\rmd t} = -\frac{\Omega}{\vA}\frac{\dd\bPhi}{\dd z},\quad
\frac{\rmd}{\rmd t} = \frac{\dd}{\dd t} + \{\bPhi,\dots\}.
\label{eq:ICW_nlin}
\eeq

In each case, the other two fields play a subordinate role: 
for KAW turbulence, from \exref{eq:Hall_Phi} and \exref{eq:Hall_U}, 
\beq
\frac{\dd}{\dd t}\dperp^2\tPhi = \vA\widetilde{\dpar\dperp^2\tPsi},
\quad
\frac{\dd\,\tU}{\dd t} = \vs\widetilde{\dpar\tB};
\label{eq:KAW_sub}
\eeq
for ICW turbulence, \exref{eq:ICW_lin_const} hold nonlinearly, viz.,
\beq
\bB = -\frac{\bPhi}{\vA\rhoh},\quad
\dperp^2\bPsi = \Omega\,\bU.
\label{eq:ICW_sub}
\eeq
The physics of these ``constitutive relations'' will be made transparent in \exref{eq:ICW_const}. 
Note that the first equation in \exref{eq:KAW_sub} combined with the second equation in 
\exref{eq:KAW_nlin} also turns into a ``constitutive relation'' 
between $\tB$ and~$\tPhi$ [cf.~\exref{eq:Hall_low_betae}]:
\beq
\dperp^2\tPhi = - \frac{\vA}{\rhoh}\,\tB.
\label{eq:KAW_sub_BPhi}
\eeq

The pieces of the free energy \exref{eq:W_Hall} 
individually conserved by the systems \exref{eq:KAW_nlin} and \exref{eq:ICW_nlin}
are, respectively,  
\begin{align}
\label{eq:WK_Hall}
\WK &= \frac{1}{2}\int\frac{\rmd^3\vr}{V}\lt[\bigl|\dperp\tPsi\bigr|^2 + \vA^2\,\tB^2\rt],\\
\WI &= \frac{1}{2}\int\frac{\rmd^3\vr}{V}\lt[\bigl|\dperp\bPhi\bigr|^2 + \vA^2\,\bU^2\rt].
\label{eq:WI_Hall}
\end{align}
Here $\WI$ is just the kinetic energy of the ion motion, perpendicular plus parallel, 
whereas $\WK$ is the total magnetic energy plus the free energy of the electron 
distribution---the latter is the $\dn^2/n^2$ term in \exref{eq:W_h}, now absorbed into 
$\tB^2$ by way of the last equation in \exref{eq:Hall_fields}.

We are now going to work out all the relevant scalings for 
KAW (\secref{sec:KAW_scalings}) and ICW (\secref{sec:ICW_scalings}) 
turbulence, then use these scalings to confirm that \exsdash{eq:KAW_nlin}{eq:ICW_sub}  
are correct (\secref{sec:Hall_decoupling}), 
and finally propose what the energy partition in these circumstances 
should be (\secref{sec:Hall_heating}). 

\subsubsection{KAW Scalings}
\label{sec:KAW_scalings}

The scalings for a critically balanced cascade of KAW-like fluctuations 
are a standard proposition (see S09--\S7.5 and \citealt{cho04}).\footnote{Various theoretical 
considerations \citep{boldyrev12,boldyrev13,meyrand13,loureiro17b,boldyrev19}, 
prompted by observational evidence \citep{alexandrova09,sahraoui10,chen16},
suggest that these ``na\"ive'' scalings may need some subtle corrections.
We shall opt for simplicity over modernity and ignore those subtleties. 
We need these scalings as a vehicle for estimating the size of KAW and ICW 
perturbations relative to each other and we do not believe that a more sophisticated 
theory of the KAW cascade will change our conclusions in any essential way.} 
The magnetic energy has a constant flux $\epsK$, with the cascade time scale
set by the magnetic nonlinearity inside $\dpar$ in, e.g., the first 
equation in \exref{eq:KAW_nlin}: 
\beq
(\kperp\tPsi)^2\tnl^{-1} \sim \epsK,\quad
\tnl^{-1}\sim\vA\rhoh\kperp^2\tB.
\label{eq:KAW_flux}
\eeq 
The relationship between $\tB$ and $\tPsi$, and hence the scaling of field amplitudes,  
is then fixed by the second equation in \exref{eq:KAW_nlin}: 
\beq
\tB \sim \oK^{-1}\rhoh\kparK\kperp^2\tPsi 
\sim \frac{\kperp\tPsi}{\vA}
\sim \lt(\frac{\epsK}{\rhoh\vA^3}\rt)^{1/3}\kperp^{-2/3},
\eeq
the last relation following from \exref{eq:KAW_flux}.
Finally, the relationship between the wave frequency $\oK$ (and, therefore, $\kparK$) 
and the nonlinear decorrelation rate $\tnl^{-1}$ (and, therefore, $\kperp$) 
is set by the critical-balance conjecture: 
\beq
\oK = \kparK\vA\kperp\rhoh \sim \tnl^{-1}
\hence
\kparK \sim \lt(\frac{\epsK}{\rhoh\vA^3}\rt)^{1/3}\kperp^{1/3}.
\label{eq:kparK}
\eeq
The two subordinate fields are found from \exref{eq:KAW_sub}: 
\beq
\tPhi \sim \oK^{-1}\kparK\vA\tPsi = \frac{\tPsi}{\kperp\rhoh},\quad
\tU \sim \oK^{-1}\kpar\vs\tB = \frac{\vs}{\vA}\frac{\tB}{\kperp\rhoh}.
\eeq
It follows from all this that the magnetic and velocity spectra are 
\beq
\label{eq:spectra_KAW}
E_{\widetilde{B}} \propto \kperp^{-7/3},\quad E_{\widetilde{u}} \propto \kperp^{-13/3} 
\eeq
\citep[cf.][]{galtier07,meyrand12}.  

\subsubsection{ICW Scalings}
\label{sec:ICW_scalings}

The scalings for a critically balanced ICW cascade are perhaps less well established, 
but also known, in the guise of the scalings for rotating hydrodynamic turbulence 
\citep{nazarenko11}. Assuming constant energy flux $\epsI$ and using the first equation 
in \exref{eq:ICW_nlin}, we find the Kolmogorov scaling (which is no surprise, the nonlinear 
coupling being hydrodynamic): 
\beq
(\kperp\bPhi)^2\tnl^{-1} \sim \epsI,\quad
\tnl^{-1} \sim \kperp^2\bPhi
\hence 
\kperp\bPhi \sim \epsI^{1/3}\kperp^{-1/3}. 
\eeq
From either equation in \exref{eq:ICW_nlin}, 
\beq
\bU \sim \frac{\kperp\bPhi}{\vA}. 
\eeq
The critical-balance conjecture implies
\beq
\oI = \Omega\,\frac{\kparI}{\kperp}\sim\tnl^{-1}
\hence 
\kparI \sim \frac{\epsI^{1/3}}{\Omega}\,\kperp^{5/3}. 
\label{eq:kparI}
\eeq
Interestingly, it follows from \exref{eq:kparI} 
that ICW turbulence becomes {\em less} anisotropic at smaller scales.\footnote{Isotropy 
is achieved at $\kperp\sim \Omega^{3/2}\epsI^{-1/2}$, known as the 
\citet{zeman94} scale in the context of inertial waves. 
This scale is, however, outside the GK ordering 
and so is formally smaller than any scale present in our considerations.} 
Finally, the subordinate fields \exref{eq:ICW_sub} are
\beq
\bB \sim \frac{\bPhi}{\vA\rhoh} \sim \frac{\bU}{\kperp\rhoh},\quad
\frac{\kperp\bPsi}{\vA} \sim \frac{\vs}{\vA}\frac{\bU}{\kperp\rhoh}.
\eeq
The velocity and magnetic energy spectra are, therefore,  
\beq
\label{eq:spectra_ICW}
E_{\overline{u}} \propto \kperp^{-5/3},\quad E_{\overline{B}}\propto \kperp^{-11/3} 
\eeq 
\citep[cf.][]{krishan04,galtier07,meyrand12}.  

\subsubsection{Decoupling of Cascades}
\label{sec:Hall_decoupling}

The above scalings appear to be consistent with the numerical evidence 
recently reported by \citet{meyrand18}, who solved the traditional Hall-MHD equations 
that effectively describe the $\beta_e\gg1$ limit of our system 
($\vs=\vA$ and $\rhoh=d_i$). They did see 
$E_{\widetilde{u}}\ll E_{\overline{B}} \ll E_{\widetilde{B}} \ll E_{\overline{u}} \propto \kperp^{-5/3}$; 
the $\kperp^{-7/3}$ and $\kperp^{-11/3}$ spectra of the magnetic perturbations associated 
with the two different wave modes [see \exref{eq:spectra_KAW} and \exref{eq:spectra_ICW}]
had previously been extracted numerically from Hall MHD by \citet{meyrand12} 
(and from a shell model by \citealt{galtier07}). 
Unlike us, \citet{meyrand18} think that the KAW turbulence is weak, 
rather than critically balanced, but 
we consider the evidence that they present in fact consistent with the possibility 
of a critically balanced KAW cascade: in particular, both their KAW fluctuations 
and their ICW fluctuations have broad frequency spectra and 
are spatially anisotropic in a scale-dependent way, the former 
becoming more anisotropic and the latter less, as $\kperp$ increases---in 
agreement with \exref{eq:kparK} and \exref{eq:kparI}. They also see striking 
evidence that $\kparK \ll \kparI$, which is indeed 
what \exref{eq:kparK} and \exref{eq:kparI} imply.  
Finally, and crucially, they show quite unambiguously that energy 
exchange between velocity and magnetic fields (and, therefore, between 
ICW and KAW fluctuations) peters out at $\kperp d_i\gg1$, i.e., 
the two cascades are energetically decoupled.  

As promised above, we now confirm that the scalings of \secsref{sec:KAW_scalings} 
and \ref{sec:ICW_scalings}, if adopted as orderings, 
do indeed allow the two cascades to decouple (an impatient reader 
willing to trust us may skip to \secref{sec:Hall_heating}). 
Let 
\beq
\epsilon = \frac{(\epsI\rhoh)^{1/3}}{\vA} \sim \frac{(\epsK\rhoh)^{1/3}}{\vA},\quad
\delta = \frac{1}{(\kperp\rhoh)^{1/3}}.
\label{eq:delta_def}
\eeq
Here $\epsilon$ is just the GK expansion parameter 
that must enter all field amplitudes. The only nontrivial choice about $\epsilon$ is that 
$\epsK\sim\epsI$, i.e., that the KAW and ICW fluctuations receive {\em a priori}
comparable amounts of energy---equivalently, we assume that the KAW and ICW amplitudes 
are similar at the Hall transition scale (at $\kperp\rhoh\sim1$). 
We now use $\delta$ as a subsidiary ordering parameter for the Hall-MHD equations 
\exsdash{eq:Hall_Psi}{eq:Hall_U}: 
\begin{align}
\label{eq:order_ICW_Phi}
&\frac{\kperp\bPhi}{\vA} \sim \bU \sim \epsilon\delta,\quad
\frac{\kperp\bPsi}{\vA}\sim \epsilon \sigma\delta^4,\quad
\bB\sim \epsilon\delta^4,\\
&\kparI\rhoh\sim \epsilon\sigma^{-1}\delta^{-5},\quad
\frac{\oI}{\Omega} \sim \epsilon\sigma^{-1}\delta^{-2},\\
&\frac{\kperp\tPsi}{\vA} \sim \tB \sim \epsilon\delta^2,\quad 
\frac{\kperp\tPhi}{\vA} \sim \epsilon\delta^5,\quad
\tU \sim \epsilon\sigma\delta^5,\\
&\kparK\rhoh \sim \epsilon\delta^{-1},\quad
\frac{\oK}{\Omega}\sim \epsilon\sigma^{-1}\delta^{-4},  
\label{eq:order_KAW_kpar}
\end{align}
where $\sigma$ is defined in \exref{eq:sigma_def}.

Applying the decomposition \exref{eq:field_split} and the above ordering to \exref{eq:Hall_Psi}, 
we get, keeping two lowest orders, 
\beq
\frac{\dd\tPsi}{\dd t} = \vA\frac{\dd}{\dd z}(\bPhi + \vA\rhoh\bB) 
+ \vA^2\rhoh\dpar\tB,\quad 
\dpar = \frac{\dd}{\dd z} + \frac{1}{\vA}\{\tPsi,\dots\}. 
\label{eq:Psi_approx}
\eeq 
Averaging this equation over the KAW time scale gives us 
\beq
\vA\frac{\dd}{\dd z}(\bPhi + \vA\rhoh\bB) 
+ \vA\rhoh\overline{\{\tPsi,\tB\}} = 0.
\label{eq:Psi_avg} 
\eeq
Subtracting this from \exref{eq:Psi_approx}, we end up with the first KAW equation in 
\exref{eq:KAW_nlin}. Retaining the lowest order only in \exref{eq:Psi_avg} 
results in the first ICW constitutive relation in \exref{eq:ICW_sub}, assuming 
that we can ignore any additive corrections to this that are constant along 
the magnetic field. 

From \exref{eq:Hall_U}, again keeping only two lowest orders, we get 
\beq
\frac{\rmd\,\bU}{\rmd t} + \frac{\dd\,\tU}{\dd t} = 
\vs\lt(\frac{\dd\bB}{\dd z} + \dpar\tB\rt),\quad 
\frac{\rmd}{\rmd t} = \frac{\dd}{\dd t} + \{\bPhi,\dots\}.
\label{eq:U_approx}
\eeq
Averaging and using \exref{eq:Psi_avg} gives us 
\beq
\frac{\rmd\,\bU}{\rmd t} = \vs\lt(\frac{\dd\bB}{\dd z} 
+ \frac{1}{\vA}\overline{\{\tPsi,\tB\}}\rt) = - \frac{\vs}{\vA\rhoh}\frac{\dd\bPhi}{\dd z},
\label{eq:U_avg}
\eeq
which is the second ICW equation in \exref{eq:ICW_nlin}. 
Subtracting \exref{eq:U_avg} from \exref{eq:U_approx} leaves us with 
the second equation in \exref{eq:KAW_sub}, describing small parallel 
ion flows associated with KAW. 

Continuing in the same vein, we find that \exref{eq:Hall_B} becomes, 
to two lowest orders, 
\beq
\frac{\dd\tB}{\dd t} = \frac{\dd}{\dd z}(\vs\bU - \rhoh\dperp^2\bPsi) 
- \rhoh\dpar\dperp^2\tPsi. 
\label{eq:B_approx}
\eeq
The average of this is
\beq
\frac{\dd}{\dd z}(\vs\bU - \rhoh\dperp^2\bPsi) 
- \frac{\rhoh}{\vA}\overline{\{\tPsi,\dperp^2\tPsi\}} = 0,
\label{eq:B_avg}
\eeq
which, to lowest order, becomes the second ICW constitutive 
relation in \exref{eq:ICW_sub} (again ignoring any contributions 
that do not vary along the magnetic field). 
Subtracting \exref{eq:B_avg} from \exref{eq:B_approx} gets us 
the second KAW equation in \exref{eq:KAW_nlin}. 

Finally, \exref{eq:Hall_Phi} to two lowest orders is
\beq
\frac{\rmd}{\rmd t}\dperp^2\bPhi + \frac{\dd}{\dd t}\dperp^2\tPhi 
= \vA\lt(\frac{\dd}{\dd z}\dperp^2\bPsi + \dpar\dperp^2\tPsi\rt). 
\label{eq:Phi_approx}
\eeq
Its average is, via \exref{eq:B_avg},
\beq
\frac{\rmd}{\rmd t}\dperp^2\bPhi = 
\vA\frac{\dd}{\dd z}\dperp^2\bPsi + \overline{\{\tPsi,\dperp^2\tPsi\}}
=\frac{\vs\vA}{\rhoh}\frac{\dd\,\bU}{\dd z},
\label{eq:Phi_avg}
\eeq
which is the first ICW equation in \exref{eq:ICW_nlin}. 
Subtracting \exref{eq:Phi_avg} from \exref{eq:Phi_approx}, 
we get the first equation in \exref{eq:KAW_sub} for the  
small perpendicular flows present in KAW. 

Thus, the equations for decoupled KAW and ICW cascades, 
\exsdash{eq:KAW_nlin}{eq:ICW_sub}, which were our basis for 
developing the scalings in \secsref{sec:KAW_scalings} and \ref{sec:ICW_scalings}, 
can indeed be extracted from Hall equations \exsdash{eq:Hall_Psi}{eq:Hall_U} 
if those scalings are assumed.\footnote{Turbulence-theory {\em literati} might 
appreciate an amusing mathematical similarity between the situation that has emerged 
here and the rigidly rotating MHD turbulence at {\em large} scales, which also 
features two co-existing cascades---of inertial and magnetostrophic waves---with 
dispersion relations and, therefore, scalings 
similar to ICW and KAW, respectively \citep{galtier14,bell19}.} 
Consistency is the least---and the most---that 
we can ask for in this approach. 

\subsubsection{Energy Partition}
\label{sec:Hall_heating}

We anticipate, and will prove in \secref{sec:Hall_Larmor}, that the sub-Hall-scale KAW 
cascade all goes into the sub-Larmor-scale KAW cascade and thence to electron heating, 
whereas the ICW cascade is destined for ion-entropy cascade and thence 
to ion heating. Thus, {\em in the Hall regime, the energy partition is decided 
at the Hall scale~$\rhoh$.} 
While we do not know how to determine this energy partition rigorously, 
a plausible conjecture can be made. 

The only parameter in the problem is 
the ratio $\sigma = \vs/\vA$ [equivalently, $\beta_e$: see \exref{eq:sigma_def}]. 
As explained at the end of \secref{sec:Hall_eqns}, 
\exsdash{eq:Hall_Psi}{eq:Hall_U} reduce to \exref{eq:RMHD_rhos} 
in the limit of $\sigma\ll1$ (low $\beta_e$). This happens because, 
sufficiently far into that limit, the finite-$\kperp\rhoh$ contribution 
to $\B$ from the Alfv\'enic fluctuations overwhelms the SW part of $\B$ 
[see \exref{eq:Hall_low_betae}], while what remains of the SW cascades 
independently according to \exref{eq:U_passive}, unbothered by the 
Hall-scale transition. The result is again \exref{eq:QiQe}: 
the Alfv\'enic energy goes into electrons, the compressive one into ions. 

In contrast, when $\sigma\sim 1$, there are no small parameters left 
in the problem and all time scales and all parts of the free energy 
\exref{eq:W_Hall} are of the same order at $\kperp\rhoh\sim1$. 
At $\kperp\rhoh\ll 1$, $\Phi$ and $\Psi$ fluctuations 
carry $\epsA$, while $\U$ and $\B$ fluctuations carry $\epsC$.
On the other side of the transition, at $\kperp\rhoh\gg 1$,
the $\Psi$ and $\B$ fluctuations (the magnetic energy) are picked up by the KAW cascade ($\epsK$)
and the $\Phi$ and $\U$ fluctuations (the kinetic energy) by the ICW cascade ($\epsI$). 
It is then natural to conjecture an equal split of the power of the original MHD cascade between 
$\epsK$ and $\epsI$, and, therefore, between electron and ion heating---independently 
of the relative size of $\epsA$ and $\epsC$:
\beq
Q_e \sim \epsK \sim \frac{\epsA + \epsC}{2} \sim \epsI \sim Q_i. 
\eeq 

Numerical simulations of the full system \exsdash{eq:Hall_Psi}{eq:Hall_U} 
with external driving are needed (and will be done) to test this reasoning.  
A parameter scan in $\sigma$ should reveal a gradual transition from 
$\lt[Q_i/Q_e\rt]_{\sigma\to0} \to \epsC/\epsA$ to 
$\lt[Q_i/Q_e\rt]_{\sigma\to1} \to 1$. 

\subsection{Helicities}
\label{sec:Hall_hel}

Before, as promised, moving on to the Larmor-scale dynamics, let us, for the 
sake of completeness and for the benefit of those readers who might be 
interested in Hall RMHD turbulence {\em per se}, offer some discussion 
of other invariants of the system \exsdash{eq:Hall_Psi}{eq:Hall_U}. 
Famously, Hall MHD equations conserve two helicities, magnetic and 
``hybrid'' \citep{turner86,mahajan98}. However, in Hall RMHD, owing 
to the presence of a strong background magnetic field, magnetic helicity 
is not conserved, except in~2D: 
\beq
H = \int\frac{\rmd^3\vr}{V}\, \Psi\B,\quad
\frac{\rmd H}{\rmd t} = \int\frac{\rmd^3\vr}{V}\,\lt(\vs\Psi\,\frac{\dd\,\U}{\dd z} 
+ \vA\B\,\frac{\dd\Phi}{\dd z}\rt) 
\eeq
(see S09--\S F.4 and references therein for a discussion of helicity non-conservation 
in a system with a mean field).
What is conserved, however, is the sum of three other ``helicities'' 
present in the system, viz., the Alfv\'enic cross-helicity, the compressive cross-helicity 
and the kinetic helicity (note that $\dperp^2\Phi$ is the $z$ component of 
the vorticity of the plasma motions):
\beq
X = \int\frac{\rmd^3\vr}{V}\,\lt[(\vdperp\Phi)\cdot(\vdperp\Psi) 
+ \frac{\vA^3}{\vs}\,\U\lt(\B + \frac{\rhoh}{\vA}\dperp^2\Phi\rt)\rt].
\label{eq:X}
\eeq   
Note that the Hall MHD ``hybrid'' helicity referred to above 
is then just $H - (\rhoh/\vA)X$ (not conserved because $H$ is not conserved). 

In the RMHD limit ($\kperp\rhoh\ll1$), $X$ loses its last term (the kinetic helicity)
and turns into the standard RMHD cross-helicity, whose 
conservation reflects the energetic decoupling of the cascades of the four Elsasser 
fields $\Phi\pm\Psi$ and $\U\pm\B$ (see S09--\S2.7). 
In the opposite limit, $\kperp\rhoh\gg 1$, 
when Hall RMHD splits into the KAW equations \exref{eq:KAW_nlin} 
and the ICW equations \exref{eq:ICW_nlin}, each of these systems conserves 
its own piece of~$X$: 
\beq
\XK = \int\frac{\rmd^3\vr}{V}\,\tPsi\tB,\quad
\XI = \frac{\vA^2}{\Omega}\int\frac{\rmd^3\vr}{V}\,\bU\dperp^2\bPhi.
\eeq
The first of these is the helicity of the KAW turbulence, which is in fact 
the cross-helicity \exref{eq:X} by way of \exref{eq:KAW_sub_BPhi} 
and integration by parts (cf.~S09--\S F.3); 
the second is the kinetic helicity of the ICW turbulence---the last 
term in \exref{eq:X}, dominant when $\kperp\rhoh\gg1$.  

While $\XK$ or $\XI$ being non-zero would indicate an imbalance 
between counterpropagating KAW-like or ICW-like perturbations, respectively, 
there is no corollary that such counterpropagating 
perturbations have energetically decoupled cascades in the way that Elsasser fields 
do in RMHD. This is because while the conserved quantity $X$ is the difference 
between the ``energies'' of the generalised 
Elsasser fields $\Phi\pm\Psi$ and $\U\pm(\B + \rhoh\dperp^2\Phi/\vA)$, 
the sum of these ``energies'' is not the free energy \exref{eq:W_Hall} 
and is not conserved, and neither, therefore, are these ``energies'' 
conserved individually. Note also that these fields are not 
the eigenfunctions associated with the counterpropagating modes: 
in the $\kperp\rhoh\gg1$ limit, those are 
$\vA\tB \mp \kperp\tPsi$ for $\omega = \pm\oK$ 
and $\vA\bU\pm\kperp\bPhi$ for $\omega = \pm\oI$. 
The energies of these fields are not individually conserved either. 

The presence of extra invariants does open the possibility 
of dual or even triple cascades in Hall RMHD turbulence: in particular, $\XK$ will cascade 
to larger scales if it is injected at small scales (see S09--\S F.6; 
\citealt{cho16} and references therein), whereas $\XI$ is expected 
to cascade forward, together with the free energy \citep{chen03,banerjee16}. 
Thus, the Hall RMHD system can offer some considerable rewards to 
a devoted turbulence theorist. 

\subsection{Larmor-Scale Transition}
\label{sec:Hall_Larmor}

We are now going to prove that, once the KAW and ICW cascades hit the Larmor 
scale, the former will be channelled into electron heating (via a sub-Larmor 
KAW cascade) and the latter into ion heating (via an electrostatic ion entropy cascade). 
What follows is formally necessary, as due diligence, but a reader who is not a particular 
GK aficionada need not read it if she trusts our algebra.\footnote{This said, 
\exref{eq:ICW_const} and \exref{eq:ICW_g} are perhaps of some technical interest, 
showing the electrostatic nature of the ICW cascade.}
Qualitative physics discussion resumes in \secref{sec:disc}. 

\subsubsection{Ordering}
\label{sec:HallLS_order}

We continue to assume the Hall ordering of the temperature ratio vs.\ plasma 
beta [see~\exref{eq:Hall_order}], 
but focus on scales that are of the order of the Larmor radius,  
a regime that does not appear to have been studied before: 
\beq
\frac{Z}{\tau} \sim \frac{1}{\beta_i}\gg1,\quad
\kperp\rho_i\sim1
\hence
\kperp\rhoh \sim \kperp\rho_i\sqrt{\frac{Z}{\tau}} \sim \frac{1}{\sqrt{\beta_i}} \gg 1. 
\label{eq:HallLS_order}
\eeq

The ordering of the time scales and amplitudes must now be adjusted. 
How to do this can be deduced {\em a priori} from the $\kperp\rhoh\gg1$ 
orderings \exsdash{eq:delta_def}{eq:order_KAW_kpar} 
by taking them to the illegitimate extreme $\kperp\rhoh\sim 1/\sqrt{\beta_i}$, 
or $\delta\sim \beta_i^{1/6}$. 
Having obtained the orderings, we will then backtrack to 
the hybrid ion-electron equations of \secsand{sec:IEF}{sec:IEF_g}
and derive a new set of equations valid under our new ordering.  

Reverting to our old notation, we convert the $\delta$ orderings 
\exsdash{eq:order_ICW_Phi}{eq:order_KAW_kpar} into $\beta_i$ orderings 
using $\delta\sim \beta_i^{1/6}$, $\sigma\sim1$, $\kpar\rhoh\sim\kpar\rho_i\sqrt{\beta_i}$, 
and 
\beq
\frac{\kperp\Phi}{\vA} \sim \frac{\kperp\rho_i\vth}{\vA}\,\ephi \sim \ephi\sqrt{\beta_i},\quad
\U \sim \frac{\upari}{\vth}\sqrt{\beta_i},\quad
\frac{\kperp\Psi}{\vA} \sim \kperp\rho_i\A \sim \A,\quad
\B \sim \frac{\dB}{B} 
\eeq
[recall \exref{eq:phi_A_def} and \exref{eq:BU_def}]. 
The resulting ordering is 
\begin{align}
&\bphi \sim \frac{\bupari}{\vth} \sim \frac{\bg}{F_0} \sim \epsilon\beta_i^{-1/3},\quad
\bA \sim \frac{\dbB}{B} \sim \frac{\dbn}{n} \sim \epsilon\beta_i^{2/3},\quad
\frac{\oI}{\Omega}\sim\frac{\kparI\vth}{\Omega}\sim\epsilon\beta_i^{-1/3},\\
&\tphi \sim \frac{\tupari}{\vth} \sim \frac{\tg}{F_0} \sim \tA \sim \frac{\dtB}{B} 
\sim \frac{\dtn}{n} \sim \epsilon\beta_i^{1/3},\quad
\frac{\oK}{\Omega}\sim\epsilon\beta_i^{-2/3},\quad
\frac{\kparK\vth}{\Omega}\sim\epsilon\beta_i^{1/3}, 
\end{align}
where bars and tildes continue to mean averaged and fluctuating quantities 
over the KAW time scale. Note that for ICW, the wave frequency and the 
ion streaming rate have turned out to be the same size, whereas for KAW, 
the former is much larger than the latter. This is the basic physical reason 
why ICW will couple into ion kinetics and, eventually, ion heating, while 
KAW will not. 

\subsubsection{Field Equations}
\label{sec:Hall_Larmor_fields}

The field equations \exsdash{eq:dn}{eq:dB} are linear, so can be split cleanly 
into slow- and fast-varying parts. 
To lowest order (in all cases, $\sim\beta_i^{-1/3}$ for the slow fields 
and $\sim\beta_i^{-2/3}$ for the fast ones), 
they are\footnote{Note that since we are now using overbars to denote 
time averages, we have suspended the overbar notation for ion velocity integrals
and reverted to writing them explicitly.} 
\begin{align}
&(1-\hG_0)\tphi = - \frac{\dtn}{n}  
+ \frac{1}{n_i}\int\rmd^3\vv\hJ_0\tg,\\
\label{eq:KAW_upare}
&\frac{\tupare}{\vth} = \frac{1}{\beta_i}\hdperp^2\tA + \tJext,\\
\label{eq:KAW_dn}
&\frac{Z}{\tau}\frac{\dtn}{n} = - \frac{2}{\beta_i}\frac{\dtB}{B},\\
\label{eq:ICW_phi}
&(1-\hG_0)\bphi = \frac{1}{n_i}\int\rmd^3\vv\hJ_0\bg,\\
\label{eq:ICW_upare}
&\frac{\bupare}{\vth} = \frac{1}{\beta_i}\hdperp^2\bA + \frac{\bupari}{\vth} + \bJext,\\
\label{eq:ICW_dn}
&\frac{Z}{\tau}\frac{\dbn}{n} = - \frac{2}{\beta_i}\frac{\dbB}{B} 
+ (1-\hG_1)\bphi - \frac{1}{n_i}\int\rmd^3\vv\hvperp^2\hJ_1\bg.
\end{align}
The external energy-injecting currents are, in fact, supposed to represent 
energy arriving from much larger scales. It is then a logical choice to 
set $\bJext=0$ and treat $\tJext$ as representing 
the incoming KAW energy [see \exref{eq:Hall_e_bal}]. 
In a similar vein, we shall, in \exref{eq:gsq_ICW}, 
let $\taext=0$ and treat $\baext$ as representing the incoming ICW energy. 

\subsubsection{Electron Equations}
\label{sec:Hall_Larmor_els}

The treatment of the electron equations \exref{eq:A} and \exref{eq:cont} 
is completely analogous to the treatment of their counterparts \exref{eq:Hall_Psi} 
and \exref{eq:Hall_B} in \secref{sec:Hall_decoupling}. We retain 
terms to two leading orders, $\beta_i^{-2/3}$ and $\beta_i^{-1/3}$:
\begin{align}
\label{eq:Hall_Larmor_A}
&\frac{\dd\tA}{\dd t} 
+ \frac{\vth}{2}\frac{\dd}{\dd z}\lt(\bphi - \frac{Z}{\tau}\frac{\dbn}{n}\rt) 
= \frac{\vth}{2}\dpar\frac{Z}{\tau}\frac{\dtn}{n} + \eta\dperp^2\tA,\\
\label{eq:Hall_Larmor_dn}
&\frac{\dd}{\dd t}\lt(\frac{\dtn}{n}-\frac{\dtB}{B}\rt) 
+ \frac{\dd\bupare}{\dd z} = - \dpar\tupare,
\rmwhere\dpar = \frac{\dd}{\dd z} - \rho_i\{\tA,\dots\}.
\end{align}
If these are averaged over the KAW time scale and then its average 
is subtracted from each equation, we obtain, using also 
\exref{eq:KAW_upare} and \exref{eq:KAW_dn}, 
\begin{align}
\label{eq:KAW_A}
&\frac{\dd\tA}{\dd t} = -\frac{\vth}{\beta_i}\widetilde{\dpar\frac{\dtB}{B}} 
+ \eta\dperp^2\tA,\\
&\frac{\dd}{\dd t}\lt(1 + \frac{2}{\beta_e}\rt)\frac{\dtB}{B} = 
\vth\widetilde{\dpar\lt(\frac{1}{\beta_i}\hdperp^2\tA + \tJext\rt)}.
\label{eq:KAW_dB}
\end{align} 
These are just the KAW equations \exref{eq:KAW_nlin} in different notation, 
but now they are valid at $\kperp\rho_i\sim1$, i.e., both above and below 
the Larmor scale. They are entirely decoupled from ion dynamics and so the 
KAW energy will cascade right through the Larmor scale and eventually 
dissipate into electron heat. 

To restate the last point in terms of a free-energy budget, the system 
\exsdash{eq:KAW_A}{eq:KAW_dB} obeys 
\beq
\label{eq:Hall_e_bal}
\frac{\rmd\WK}{\rmd t} + Q_e = 
\vth^2 \int\frac{\rmd^3\vr}{V}\frac{\dd\tA}{\dd t} \tJext = \epsK, 
\eeq 
where $\WK$ is given by \exref{eq:WK_Hall} [it is the same as the last 
two terms of \exref{eq:W_g}, after using \exref{eq:KAW_dn}], 
$Q_e$ is given by \exref{eq:Qe_def} (with $\A\to\tA$) 
and $\tJext$ now represents the KAW cascade from $\kperp\rho_i\ll1$. 
In steady state, 
\beq
Q_e = \epsK.
\eeq

Returning to the averaged versions 
of \exref{eq:Hall_Larmor_A} and \exref{eq:Hall_Larmor_dn}
and retaining only the lowest order, we find 
\beq
\frac{\dd}{\dd z}\lt(\bphi - \frac{Z}{\tau}\frac{\dbn}{n}\rt) = 0,\quad
\frac{\dd\bupare}{\dd z} = 0. 
\label{eq:ICW_const}
\eeq
With the aid of \exref{eq:ICW_upare} and \exref{eq:ICW_dn}, 
these are readily seen to be the $\kperp\rho_i\sim1$ counterparts of 
the ICW ``constitutive relations'' \exref{eq:ICW_sub}. 
When they are written in the form \exref{eq:ICW_const}, their 
physical meaning becomes particularly transparent: these are 
statements of Boltzmann (``adiabatic'') electrons and zero electron current, 
usually associated with the electrostatic approximation. 
We shall see in \secref{sec:Hall_Larmor_ions} that ion dynamics 
on ICW time scales are indeed electrostatic. 

\subsubsection{Ion Equations}
\label{sec:Hall_Larmor_ions}

Finally, we treat the ion GK equation \exref{eq:g} in the same 
manner as we did the electron equations in \secref{sec:Hall_Larmor_els}. 
To two lowest orders, it is 
\begin{align}
\nonumber
\frac{\dd\bg}{\dd t} &+ \frac{\rho_i\vth}{2}\{\la\bphi\ra_\vR,\bg\}
+ \frac{\dd}{\dd t}\lt(\tg - \hvperp^2\hJ_1\frac{\dtB}{B}F_0\rt) 
+ \vpar\lt\la\frac{\dd}{\dd z}\lt(\bg + \frac{Z}{\tau}\frac{\dbn}{n}F_0\rt)
+ \dpar\frac{Z}{\tau}\frac{\dtn}{n}F_0\rt\ra_\vR\\ 
&= C[\bg + \tg + \la\bphi + \tphi\ra_\vR F_0] 
+ \frac{2\vpar\la\baext + \taext\ra_\vR}{\vth^2}F_0.
\label{eq:Hall_Larmor_g}
\end{align}
Taking the KAW-time-scale average of \exref{eq:Hall_Larmor_g} and then 
subtracting it from the equation, we~get 
\beq
\frac{\dd}{\dd t}\lt(\tg - \hvperp^2\hJ_1\frac{\dtB}{B}F_0\rt) 
= \vpar\lt\la\widetilde{\dpar\frac{2}{\beta_i}\frac{\dtB}{B}}\rt\ra_\vR F_0
+ C[\tg + \la\tphi\ra_\vR F_0], 
\label{eq:KAW_g}
\eeq
where have used also \exref{eq:KAW_dn} and set $\taext=0$ as promised 
at the end of \secref{sec:Hall_Larmor_fields}. This is an (irrelevant) 
imprint of the KAW turbulence on the ion distribution function---the 
FLR version of \exref{eq:KAW_sub}.\footnote{The first and second equations 
of \exref{eq:KAW_sub} are recovered by taking the density and parallel-velocity 
moments, respectively, of \exref{eq:KAW_g}, using, in the case of the density 
moment, \exref{eq:KAW_dB}, and going to the $\kperp\rho_i\ll1$ limit.}  
Note that there is no phase mixing 
here, either parallel or perpendicular, so, in a weakly collisional 
plasma, these small perturbations of the ion distribution function 
have no means of accessing the collision operator and thermalising. 

Returning to the KAW-time-scale average of \exref{eq:Hall_Larmor_g}, 
retaining the lowest-order terms and using the first equation in 
\exref{eq:ICW_const}, we get 
\beq
\frac{\dd\bg}{\dd t} + \frac{\rho_i\vth}{2}\{\la\bphi\ra_\vR,\bg\}
+ \vpar\frac{\dd}{\dd z}\lt(\bg + \la\bphi\ra_\vR F_0\rt) = 
C[\bg + \la\bphi\ra_\vR F_0] + \frac{2\vpar\la\baext\ra_\vR}{\vth^2}F_0. 
\label{eq:ICW_g}
\eeq 
Together with \exref{eq:ICW_phi}, this is a closed system---the 
standard electrostatic GK equation supporting ion hydrodynamics 
\exref{eq:ICW_nlin} at long scales ($\kperp\rho_i\ll1$)\footnote{This is again derived 
by taking the density and parallel-velocity moments of \exref{eq:ICW_g}.} 
and the ion entropy cascade at sub-Larmor scales (see S09--\S7.10 and \citealt{sch08}). 
There is no coupling to any other dynamics and so the ICW 
energy arriving from $\kperp\rho_i\ll1$ flows into the ion 
entropy cascade at $\kperp\rho_i\gg1$, to become, upon reaching 
collisional phase-space scales, ion heat. 

For the reference of a meticulous reader, the dispersion relation 
that follows from \exref{eq:ICW_g} and \exref{eq:ICW_phi}~is
\beq
1 + \zeta\Z(\zeta) = 1 - \frac{1}{\Gamma_0},\quad
\zeta = \frac{\omega}{|\kpar|\vth},
\label{eq:ICW_DR}
\eeq
where $\Z(\zeta)$ is the plasma dispersion function \citep{fried61}.
When $\kperp\rho_i\ll1$ and (consequently) $\zeta\gg1$, 
the right-hand side of \exref{eq:ICW_DR} 
is $\approx - \kperp^2\rho_i^2/2$ and the left-hand side 
is $\approx-1/2\zeta^2$ (the ``fluid'' limit). 
The result is the ICW dispersion relation~\exref{eq:ICW}. 
When $\kperp\rho_i\sim1$, we must have $\zeta\sim1$ and the solutions 
of \exref{eq:ICW_DR} contain heavy Landau damping on the ions---the linear signature 
of ion heating. 

Finally, the free-energy budget of the system \exref{eq:ICW_g} and \exref{eq:ICW_phi}~is 
\beq
\frac{\rmd}{\rmd t}\frac{\vth^2}{4}\!\int\!\frac{\rmd^3\vr}{V}\lt[
\frac{1}{n_i}\!\int\!\rmd^3\vv\frac{\la \bg^2\ra_\vr}{F_0} 
+ \bphi(1-\hG_0)\bphi\rt] + Q_i 
= \int\!\frac{\rmd^3\vr}{V}\frac{1}{n_i}\!\int\!\rmd^3\vv\,
\baext\la\vpar g\ra_\vr = \epsI, 
\label{eq:gsq_ICW}
\eeq
where $Q_i$ is given by \exref{eq:Qi_def} (with $h\to\bh$) 
and, as promised in \secref{sec:Hall_Larmor_fields}, $\baext$ now represents 
the energy flux into the ICW cascade. 
In the long-wavelength limit $\kperp\rho_i\ll1$, 
the individually conserved piece of free energy appearing in the left-hand side 
turns into the ICW free energy \exref{eq:WI_Hall} plus the variance of the 
passive kinetic field $\G$ [see \exref{eq:Hall_G}]. The difference between 
\exref{eq:gsq_ICW} and the analogous low-$\beta_e$  equation \exref{eq:gsq_ev} 
is that the ``kinetic-energy'' term $\bphi(1-\hG_0)\bphi$ has now migrated 
into the ion-heating balance [cf.~\exref{eq:W_coupled} and \exref{eq:W_g}] 
(removing also the technical complications associated with $\Qx$).  
In steady state, \exref{eq:gsq_ICW} tells us that 
\beq
Q_i = \epsI,
\eeq
restating again that all the ICW energy goes into ion heating. 

\section{Discussion}
\label{sec:disc}

The physics of turbulent heating of low-beta GK plasmas was already summarised 
and discussed at length in \secref{sec:epitome}, so we need not repeat that discussion. 
The headline result is the clean separation between the Alfv\'enic cascade heating the 
electrons and the compressive cascade the ions, at low $\beta_i$ and low $\beta_e$ 
(\secref{sec:low_betae}). One practical implication is that it becomes 
an interesting question, not just in itself, but also for large-scale modelling 
of, e.g., detectable emission from astrophysical objects 
\citep[e.g.,][]{ressler17,chael18,chael19}, 
how any particular type of MHD turbulence present in these objects splits itself 
into these two cascades---the answer to this question for, e.g., MRI turbulence, 
is not known, although it can, in principle, be obtained via standard fluid simulations.  
In the solar wind, the answer is known observationally, if not necessarily understood 
theoretically: the compressive cascade carries about $10\%$ of the energy
\citep{howes12,chen16}. 

Obviously, it must be appreciated that our prediction for the energy partition 
is only as good as the GK (low-frequency) approximation that has 
been used to make it. The most developed theoretical scheme that breaks this approximation 
and provides some significant ion heating is the so called ``stochastic heating'', 
caused by turbulent fluctuations distorting ions' Larmor orbits \citep{chandran10}; 
other possibilities involve various forms of cyclotron heating of 
the ions \citep[e.g.][]{gary05,kasper08,kasper13,marsch11,arzamasskiy19}.  
Thus, our prediction of ion heating should perhaps be viewed as a lower bound. 

Let us discuss very briefly the conditions under which the stochastic heating might 
take over (a more sophisticated recent take on this topic can be found in \citealt{mallet18}). 

\subsection{Stochastic Heating}
\label{sec:sh}

The fraction of the Alfv\'enic energy flux arriving to the ion Larmor scale that 
gets converted into ion heat via the stochastic heating mechanism is \citep{chandran10}
\beq
Q_i^\mathrm{(stoch)} \sim \epsA e^{-1/\delta},\quad
\delta \sim \frac{u_{\perp\rho_i}}{\vthi} 
\sim \frac{1}{\sqrt{\beta_i}}\frac{\dB_{\perp\rho_i}}{B_0}
\sim \frac{1}{\sqrt{\beta_i}}\frac{\dB_{\perp L}}{B_0}\lt(\frac{\rho_i}{L}\rt)^{1/3},
\label{eq:stoch_heating}
\eeq
where $u_{\perp,\rho_i}$ and $\dB_{\perp\rho_i}$ are the typical velocity and magnetic perturbations 
at the Larmor scale. The last estimate comes from assuming a $\kperp^{-5/3}$ spectrum 
of the Alfv\'enic cascade (replace the exponent $1/3$ with $1/4$ if you prefer $\kperp^{-3/2}$), 
to refer $\delta$ to the magnetic-field variation $\dB_{\perp L}$ at the outer scale $L$. 
Given $L$ and $\dB_{\perp L}$, which are independent, system-specific properties,  
setting $\delta\sim1$ in \exref{eq:stoch_heating} gives us an estimate of the limitations 
of both the GK and low-beta limits: indeed, in the ordering of 
\secref{sec:low_beta_ordering}, $\delta\sim\epsilon/\sqrt{\beta_i}$, so 
$\delta\sim1$ is when these two limits clash. 
In the solar wind, usually, $\dB_{\perp L}/B_0\sim1$ and $\rho_i/L\sim 10^{-4}$, 
so, if we were to err on the side of caution, we would start disbelieving 
the GK predictions for $\beta_i\lesssim10^{-2}$, although it is not 
hard to play with numbers and lower this by another factor of 10 in specific 
circumstances. More careful estimates of the validity of the GK approximation
can be found in \citet{howes08jgr} and of the importance of stochastic heating 
in \citet{chandran10} and \citet{chandran10corona}. 
Our purpose here is to emphasise that the constraints 
that we have placed on the ion heating are pessimistic (from the ions' viewpoint) 
and may become unreliable when $\beta_i$ is too low.\footnote{Note the recent observational 
analysis by \citet{vech17} and theoretical arguments by \citet{mallet18} 
and \citet{hoppock18}, which suggest 
that stochastic heating may, quantitatively, be more important, 
at higher values of $\beta_i$, than previously believed.}  

An interesting corollary is that there might be an intrinsic mechanism that would prevent 
$\beta_i$ from being much lower than the stochastic-heating threshold: 
indeed, if $\beta_i$ did drop lower, stochastic heating would become significant 
and channel turbulent energy into ions, which would increase 
$\beta_i$ and shut down stochastic heating. One could imagine some equilibrium 
hovering around that threshold in a system where ions, starved of heating 
in the GK approximation, were able to cool down and thus lower $\beta_i$ 
until stochastic heating turned on.\\ 

It is perhaps useful to mention two other plausible self-regulation mechanisms 
implied by our considerations above.   

\subsection{Energy Redistribution in the Hall Regime} 

At the price of the rather long derivation in \secref{sec:Hall}, we learned 
that the clean energy partition that holds at low $\beta_e$ breaks 
down at $\beta_e\gtrsim 1$. If, in a given low-$\beta_i$ system, 
electrons are heated preferentially and if that preferential heating leads 
to electron temperature increasing, then the system will be nudged towards 
the Hall limit.\footnote{If $T_e$ does not change, changing $T_i$ alone 
cannot, obviously, alter the relative size of $\tau=T_i/T_e$ and $\beta_i$ 
because both parameters are~$\propto T_i$.} 
Once $T_i/T_e \sim \beta_i$ (equivalently, $\beta_e\sim1$),
electrons will have to start 
sharing turbulent energy with ions, probably equally (\secref{sec:Hall_heating}). 
This may mean, effectively, that $T_i/T_e$ cannot decrease further 
and/or that $\beta_i$ will be pushed up. Thus, low-$\beta_i$ plasma is 
intrinsically averse to electrons getting too hot. 
 
\subsection{Collisional Heating} 

In all of the above (and, in particular, in \secref{sec:low_beta_partition}), 
we have assumed that ion collisions are sufficiently infrequent 
for the collision operator to become important only at sub-Larmor scales. 
If, however, ions are starved of heating and are, as a result, cooled by some 
competing mechanism, their collision frequency will increase. The typical rate 
at which collisional heating happens is [cf.~\exref{eq:Qi_est}] 
$\tcoll^{-1}\sim\nu_{ii}(\kperp\rho_i)^2$. This is to be compared with 
the turbulent-cascade rate: for Alfv\'enic turbulence, 
$\tnl^{-1}\sim\kperp\uperp\sim \epsA^{1/3}\kperp^{2/3}$. 
Balancing the two rates gives us a ``Kolmogorov scale''
\beq
\tcoll^{-1} \sim \tnl^{-1} 
\hence 
\kcoll\rho_i \sim \frac{\epsA^{1/4}}{\rho_i^{1/2}\nu_{ii}^{3/4}} 
\propto \epsA^{1/4}n_i^{-3/4}B^{1/2}T_i^{7/8}. 
\eeq
If $T_i$ is so low that $\kcoll\rho_i\lesssim1$, the cascade will 
be dissipated by ion (perpendicular) viscosity and ion heating will result. 
Again, one can imagine an equilibrium hovering around the transition 
between the two regimes---collisional and collisionless.\\ 

While it is not our purpose here to propose macroscopic thermodynamic models of any
specific object, 
we hope that we have given a more object-oriented reader some food for thought 
and perhaps even some useful information, while a fellow kinetic-theory enthusiast 
might have enjoyed the ride. Some of the ideas, loose ends and 
opportunities for numerical verification identified 
above will be picked up in our own future work. 

\begin{acknowledgments} 
We are indebted to W.~Dorland, T.~Adkins, S.~Balbus, S.~Cowley, 
N.~Loureiro, F.~Parra, and E.~Quataert for many important conversations,  
to B.~Chandran, G.~Howes and M.~Kunz for detailed comments on the draft manuscript,  
and to R.~Meyrand for a tutorial on Hall MHD 
The work of YK was supported in full and of MAB and AAS 
in part by the UK STFC Consolidated Grant ST/N000919/1; 
AAS was also supported in part by the UK EPSRC Grant EP/M022331/1 
and, at NBIA, by the Simons Foundation. 
The authors are grateful to the Wolfgang Pauli Institute, Vienna, 
for its hospitality and the scientific interactions that it enabled, on multiple occasions. 
\end{acknowledgments} 

\bibliography{../../JPP/bib_JPP}{}
\bibliographystyle{jpp}

\end{document}